\documentclass{elsart}
\usepackage{graphicx}
\usepackage{epsfig}
\usepackage{amssymb}
\usepackage{color}

\begin{document}
\begin{frontmatter}

\title{The ground state energy of the Edwards-Anderson spin glass model
with a parallel tempering Monte Carlo algorithm}

\author{F. Rom\'a$^{1,2}$},
\author{S. Risau-Gusman$^1$},
\author{A. J. Ramirez-Pastor$^2$},
\author{F. Nieto$^2$},
\author{E. E. Vogel$^3$}


\address{ $^1$ Centro At{\'{o}}mico Bariloche, San
Carlos de Bariloche, R\'{\i}o Negro R8402AGP , Argentina
\\ $^2$ Universidad Nacional de San Luis, Chacabuco 917, San Luis D5700BWS, Argentina
\\ $^3$ Departamento de F\'{\i}sica, Universidad de La
Frontera, Casilla 54-D, Temuco, Chile}

\begin{abstract}
We study the efficiency of parallel tempering Monte Carlo
technique for calculating true ground states of the
Edwards-Anderson spin glass model. Bimodal and Gaussian bond
distributions were considered in two and three-dimensional
lattices. By a systematic analysis we find a simple formula to
estimate the values of the parameters needed in the algorithm to
find the GS with a fixed average probability. We also study the
performance of the algorithm for single samples, quantifying the
difference between samples where the GS is hard, or easy, to find.
The GS energies we obtain are in good agreement with the values
found in the literature. Our results show that the performance of
the parallel tempering technique is comparable to more powerful
heuristics developed to find the ground state of Ising spin glass
systems.
\end{abstract}




\end{frontmatter}

\section{Introduction}
The study of spin glasses is an active and controversial area of
statistical physics. In particular, the properties of these
systems at zero temperature have been intensively studied in the
last years. The problem of finding ground states (GSs) is a very
difficult subject because of the quenched disorder and frustration
that are present in most realistic spin glass models. In fact, it
has even been shown that finding the GS of a spin glass in a
three-dimensional lattice is an NP-complete problem
\cite{Barahona1982}, which means that this challenge is at least
as difficult as the hardest problems of practical interest. As a
consequence, many different algorithms
\cite{Holland1975,Pal1995,Pal1996a,Hartmann1996,Hartmann1997,Houdayer2001,Alder2004}
have been proposed to solve it, and to assess the efficiency of
these algorithms is thus very important. {\it Genetic algorithms}
\cite{Holland1975} are considered as the most powerful heuristics
to reach the GS of spin glass systems
\cite{Hartmannlibro1,Hartmannlibro2}. After introduction of the
triadic crossover by P\'al \cite{Pal1995,Pal1996a}, different
improvements have been made to combine genetic algorithms with,
for example, {\it cluster-exact approximation}
\cite{Hartmann1996,Hartmann1997} and {\it renormalization}
\cite{Houdayer2001}. Nevertheless, algorithms based in Monte Carlo
(MC) methods, such as {\it simulated annealing} (SA)
\cite{Kirkpatrick1983}, {\it multicanonical ensemble}
\cite{Berg1994} and {\it parallel tempering} (PT)
\cite{Geyer1991,Hukushima1996}, have also been used to find GSs of
spin glasses. Just as genetic algorithms, they are simple to be
implemented. However, MC methods are usually considered less
efficient than genetic algorithms, because it is often assumed
that the presence of a temperature parameter in the algorithm
entails a breaking of the ergodicity that can lead to difficulties
in the searching of GSs of disordered systems.

Recently, PT has been used to find the GS (see, for instance
Refs.~\cite{Moreno2003,Katzgraber2003a,Katzgraber2003b,Roma2007,Risau2008}),
and it has been shown \cite{Moreno2003} to be more efficient than
other MC based algorithms. However, the issue of the efficiency of
the PT algorithm has been very briefly discussed in the
literature. In this work, we tackle this point in a more
systematic way by analyzing how this efficiency depends on the
different input parameters of the PT. The system to which the
algorithm is applied is the Edwards-Anderson (EA) spin glass model
\cite{Edwards1975} in two-dimensional (2D) and three-dimensional
(3D) lattices, with both bimodal and Gaussian distributions of
bonds. The results show that the performance of the PT technique
is comparable to more powerful heuristics developed to find the GS
of disordered and frustrated systems. Furthermore, we show that
the efficiency depends of certain combinations of the parameters.
In particular, we find a heuristic formula that gives the minimum
time (in unit of time of PT algorithm, see below) that is
necessary to find the GS with a fixed probability and for a given
lattice size.

The paper is structured as follows.  In Sec.~\ref{sec2} we present
the EA model and three different implementations of the PT
algorithm. In Sec.~\ref{sec3}, we determine the values of the
algorithm parameters that are needed to find the GS, with a fixed
{\it average} probability, for 2D and 3D EA models with bimodal
and Gaussian bond distributions, for small lattice sizes. Using
this, we obtain the GS energy for larger sizes and the
thermodynamic limit of this quantity. In Sec.~\ref{sec4}, we study
the efficiency of the PT algorithm to find the GS of {\it single}
samples. Conclusions are drawn in Sec.~\ref{sec5}.

\section{Model and Algorithm \label{sec2}}
We consider the Edwards-Anderson spin glass model
\cite{Edwards1975}, which consists of a set of $N$ Ising spins
$\sigma_i = \pm 1$ placed in a square or cubic lattice of linear
dimension $L$, with periodic boundary conditions in all
directions. Its Hamiltonian is
\begin{equation}
H = \sum_{( i,j )} J_{ij} \sigma_{i} \sigma_{j},
\label{hamiltonian}
\end{equation}
\noindent where $( i,j )$ indicates a sum over nearest neighbors.
The coupling constants or bonds, $J_{ij}$'s, are independent
random variables drawn from a given distribution with mean zero
and variance one. In this paper we concentrate on the EA model
with two distributions: the bimodal (EAB), and the Gaussian (EAG).
In the EAB model, the bonds take only two values $J_{ij}=\pm 1$,
with equal probability. For relatively large system sizes, and due
to the fact that the bonds are independent variables, only
configurations with half of the bonds of each sign are
statistically significant. To preserve this feature for small
sizes, we explicitly enforce the constraint
\begin{equation}
\sum_{( i,j )} J_{ij} = \left\{
\begin{array}{cl}
0 & \mbox{for even number of bonds} \\
\pm 1 & \mbox{for odd number of bonds}.
\end{array}
\right.
\label{constraint}
\end{equation}
For systems with an odd number of bonds, we enforce the constraint
$\sum_{( i,j )} J_{ij} =1$ for the half of the samples and
$\sum_{( i,j )} J_{ij} =-1$ for the other half. In the EAG model,
the bonds are drawn from a Gaussian distribution. One important
difference between these two models is that whereas for the EAG
the GS of the system is unique (up to a global symmetry), the EAB
has a highly degenerate GS.

In order to implement a PT algorithm \cite{Hukushima1996} one
needs to make $m$ replicas of the system (ensemble) to be
analyzed, each of which is characterized by a temperature
parameter $T_i$ ($T_1 \geq T_i \geq T_m$). The basic idea of this
algorithm is to simulate independently a Hamiltonian dynamics
(standard MC) for each replica, and to swap periodically the
configurations of two randomly chosen temperatures. The purpose of
this swap is to try to avoid that replicas at low temperatures get
stuck in local minima. Thus the highest temperature, $T_1$, is set
in the high-temperature phase where relaxation time is expected to
be very short and there exists only one minimum in the free energy
space. The lowest temperature, $T_m$, is set in the
low-temperature phase. Within this interval we choose equally
spaced temperatures, i.e. $T_i-T_{i+1}= \left(T_{1} - T_m
\right)/(m-1)$.

As mentioned above, PT is based on two procedures that are
performed alternately. In the first one, a standard MC method is
used to independently simulate the dynamics of each replica: in
each elementary step, the update of a randomly selected spin of
the ensemble is attempted with a probability given by the
Metropolis rule \cite{Metropolis}. In the second procedure, a
trial exchange of two configurations $X_i$ and $X_{i'}$
(corresponding to the $i$-th and $i'$-th replicas) is attempted,
and accepted with probability \cite{Hukushima1996}
\begin{equation}
W\left(X_i,\beta_i| X_{i'},\beta_{i'}\right)=\left\{
\begin{array}{cc}
1 & {\rm for}\ \ {\Delta \leq 0} \\
\exp(-\Delta)  & {\rm for}\ \ {\Delta>0},
\end{array}
\right.\label{exchange}
\end{equation}
where $\Delta=\left(\beta_{i'} -  \beta_i \right)\left[ H(X_i) -
H(X_{i'}) \right]$ and $\beta_i = 1/T_i$ (we have taken the
Boltzmann's constant equal to one without loss of generality). As
in Ref.~\cite{Hukushima1996}, we restrict the replica exchange to
the case $i'= i+1$.  The unit of time in this process or PT step
(PTS) consists of a fixed number of elementary steps of standard
MC, followed by other fixed number of trials of replica exchange.
The initial configuration of each replica is usually random but,
as discussed below, other choices can endow the algorithm with
some interesting features. The running time of our code,
$t_{\mathrm{sec}}$ (in seconds), can be approximated by
\begin{equation}
t_{\mathrm{sec}}=\alpha ~m ~n ~N ~t , \label{tsec}
\end{equation}
where $t$ is the number of PTSs, $n$ is the number of independent
runs and $\alpha$ is a constant.

Depending on the way we combine the number of elementary steps of
standard MC and the number of replica exchanges, we define three
different variants of the PT algorithm. The one that we call A
algorithm consists of two stages. The first is simply a SA routine
implemented as follows. Starting from a random initial condition,
$t_{\mathrm{A}} /2$ MC steps (MCSs) of standard MC are applied to
replica 1 (each MCS consists of $N$ elementary steps of standard
MC). Next, the same is done successively for each replica, but
taking the last configuration of replica $i$ as the initial
condition of replica $i+1$. The final configurations obtained are
used to initialize a PT algorithm. In the second stage, we define
that a PTS consists of $m\times N$ cycles, each cycle being one
elementary step of standard MC plus one replica exchange. After
$t_{\mathrm{A}}/2$ PTSs, the algorithm stops and its output is the
configuration with the smallest energy among all configurations
visited by all replicas in the simulation process.

In the second variant, that we call B, the $m$ replicas are
initialized with a random configuration and the PTS consists of
$m\times N$ elementary steps of standard MC and only one replica
exchange. This definition is usually chosen to reach equilibrium.
After $t_{\mathrm{B}}$ PTSs the algorithm stops and the
configuration with lowest energy is stored.

Finally, a third variant, called C, consists only of the second
stage of variant A, but with the initial configurations randomly
chosen. After $t_{\mathrm{C}}$ PTSs (where the PTS is defined as
in A algorithm), the configuration with minimum energy is stored.

To compare the performance of the three variants proposed, we run
each one of them on the same $N_{\mathrm{s}}=10^3$ samples of the
2D EAB model (only one run for sample). We choose the values of
parameters $t_{\mathrm{A}}$, $t_{\mathrm{B}}$ and $t_{\mathrm{C}}$
in such a way as to ensure that the running time of the three
variants is the same. In our case, we have used
$t_{\mathrm{A}}=2t$, $t_{\mathrm{B}}=2.3t$ and
$t_{\mathrm{C}}=1.5t$ (this choice depends on the particular
implementation of each algorithm).  In order to check whether the
final configuration found by each algorithm is really a GS, we
compare with the output of an exact branch-and-cut algorithm run
on the same samples \cite{DeSimone1995,Cologne}. The quantity we
choose to compare the efficiency of the variants A, B and C, is
the mean probability of finding the GS, $\mathcal{P}_0$, which we
estimate as
\begin{equation}
P_0=\frac{1}{N_{\mathrm{s}}}\sum_{j=1}^{N_{\mathrm{s}}}P_{0,j},
 \label{P0}
\end{equation}
where
\begin{equation}
P_{0,j}=\frac{n_j}{n},
 \label{P0j}
\end{equation}
is an estimation of the probability of reaching the GS for the
$j$-th sample, $\mathcal{P}_{0,j}$. In the last equation, $n_j$ is
the number of times that GS is found for the $j$-th sample in $n$
independent runs.  Note that in this example we use $n=1$.
Therefore for each sample $P_{0,j}=0$ or 1. As it is shown in the
appendix A, the error associated to $P_0$  becomes small if many
samples are considered and only one run is carried out in each one
of them (it is not necessary to consider many runs for sample,
i.e. $n>>1$).

\begin{figure} [t]
\includegraphics[width=8cm,clip=true]{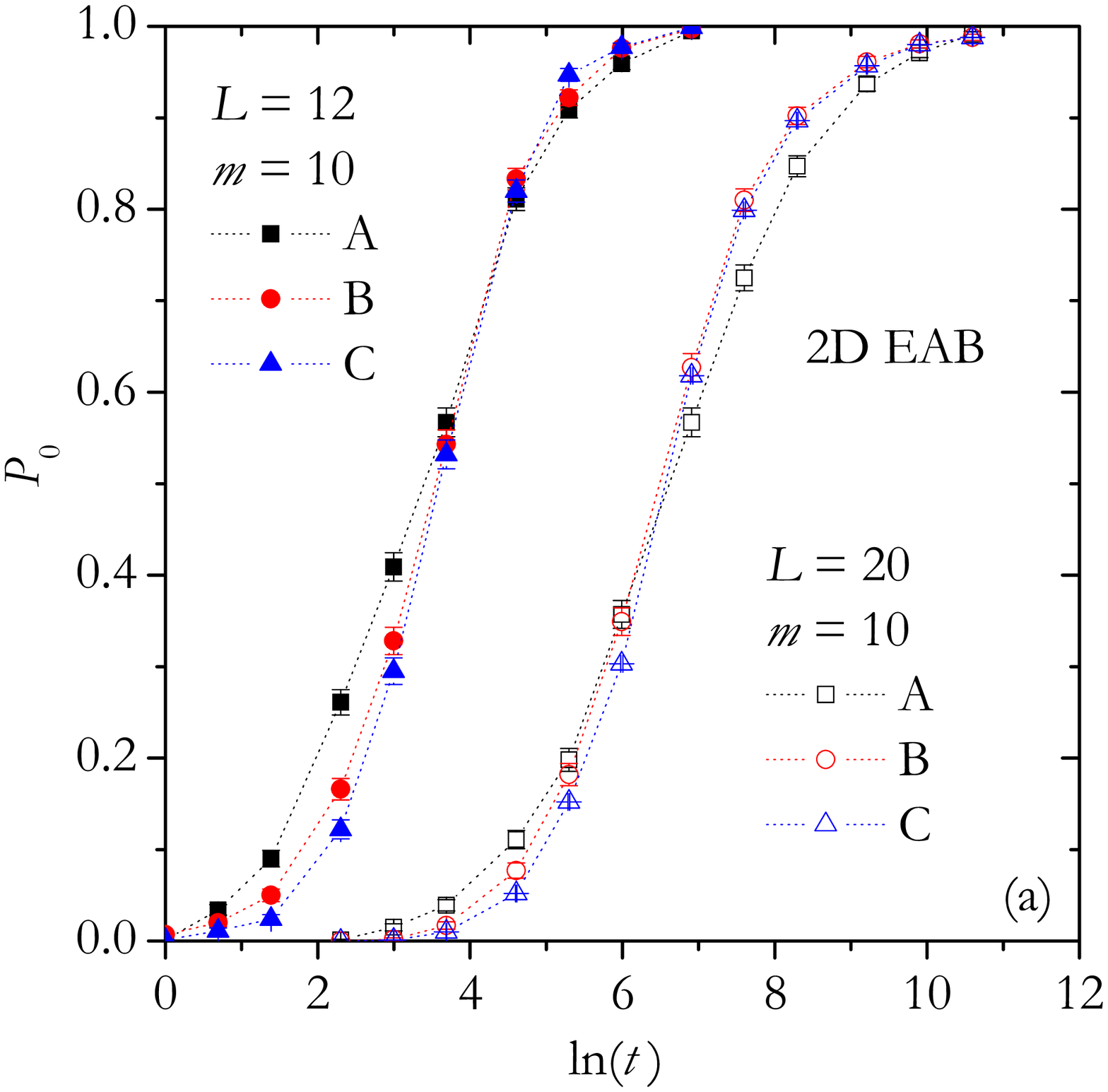}
\includegraphics[width=8cm,clip=true]{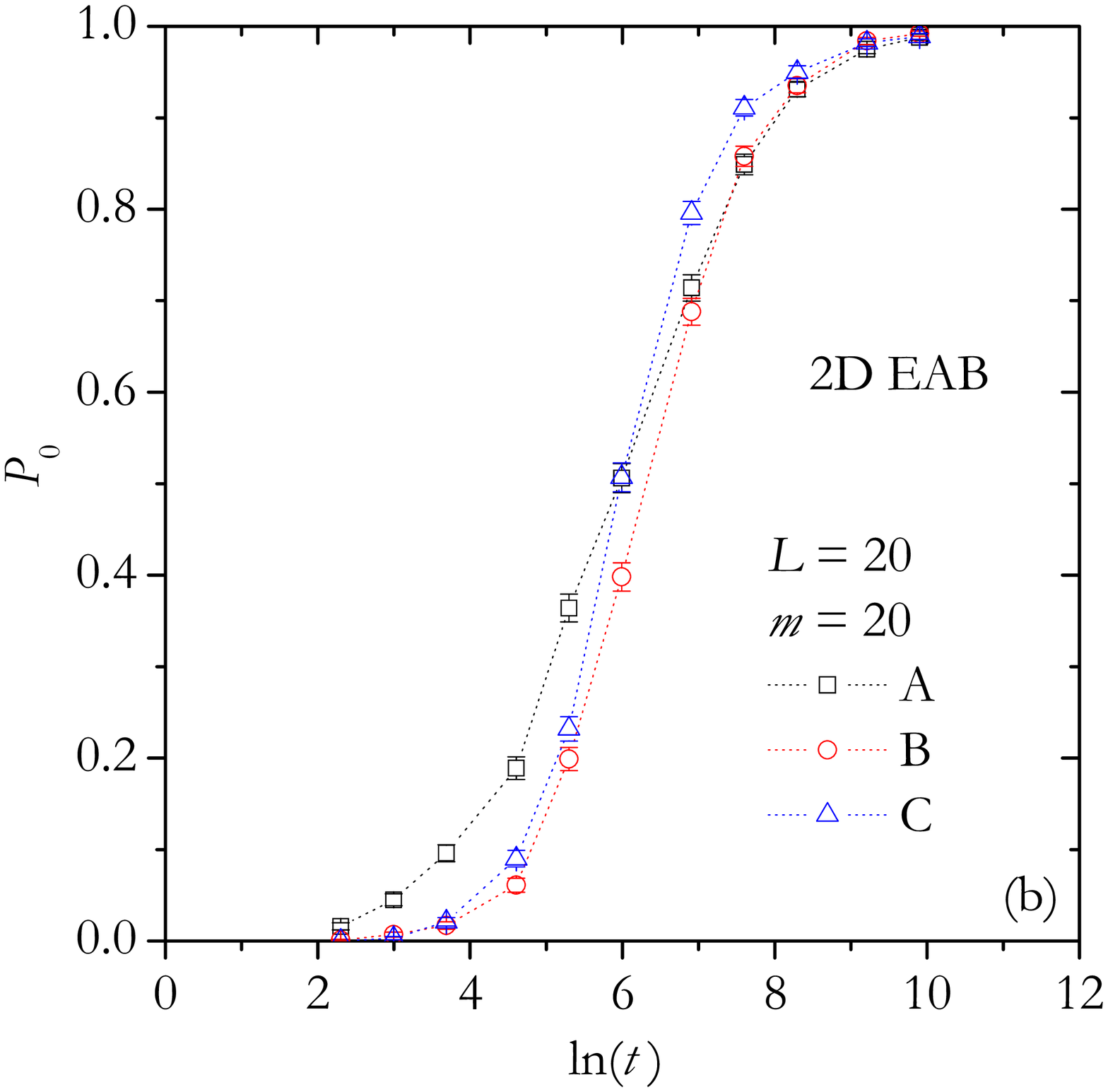}
\caption{\label{figure1}  $P_0$ as function of $\ln (t)$ for the
2D EAB model, calculated with three variants of the PT algorithm.
In all cases, the set of temperatures used varies between
$T_1=1.6$ and $T_m=0.1$. (a) $L=12$ and $L=20$ with $m=10$, and
(b) $L=20$ with $m=20$.}
\end{figure}

The result of this comparison is shown in Fig.~\ref{figure1} for
different values of $L$ and $m$ (see appendix A for a detailed
calculation of the error bars). It can be seen that the
performance of the three variants is very similar for all values
of $t$. However, a exhaustive analysis shows that for $P_0 > 0.5$,
the performance of B and C algorithms is a little better than the
one corresponding to A and this behavior is reinforced upon
increasing the lattice size [see Fig.~\ref{figure1} (a)]. On the
other hand, by comparing the curves for $L=20$ in
Figs.~\ref{figure1} (a) and (b), we observe that if $m$ is
increased, the probability $P_0$ for A and B is the same, while
for C it is a little larger. These examples show that the PT
algorithm has a complex dependence with $m$. The above comparison
has been performed also for the other studied cases (2D EAG, 3D
EAB and 3D EAG models) and the results are very similar to those
shown in Fig.~\ref{figure1}. The most important feature is that in
the large $P_0$ regime, the performance of A is always worse than
the other two, which means that its performance could be used as a
lower bound for the B and C cases. For this reason, in the
following we only analyze the performance of A. In addition, we
show below that variant A presents very interesting scaling
properties. To avoid confusions, we keep $t_{\mathrm{A}}=2t$ in
the rest of this paper.

\section{Results \label{sec3}}

\begin{figure}[t]
\includegraphics[width=8cm,clip=true]{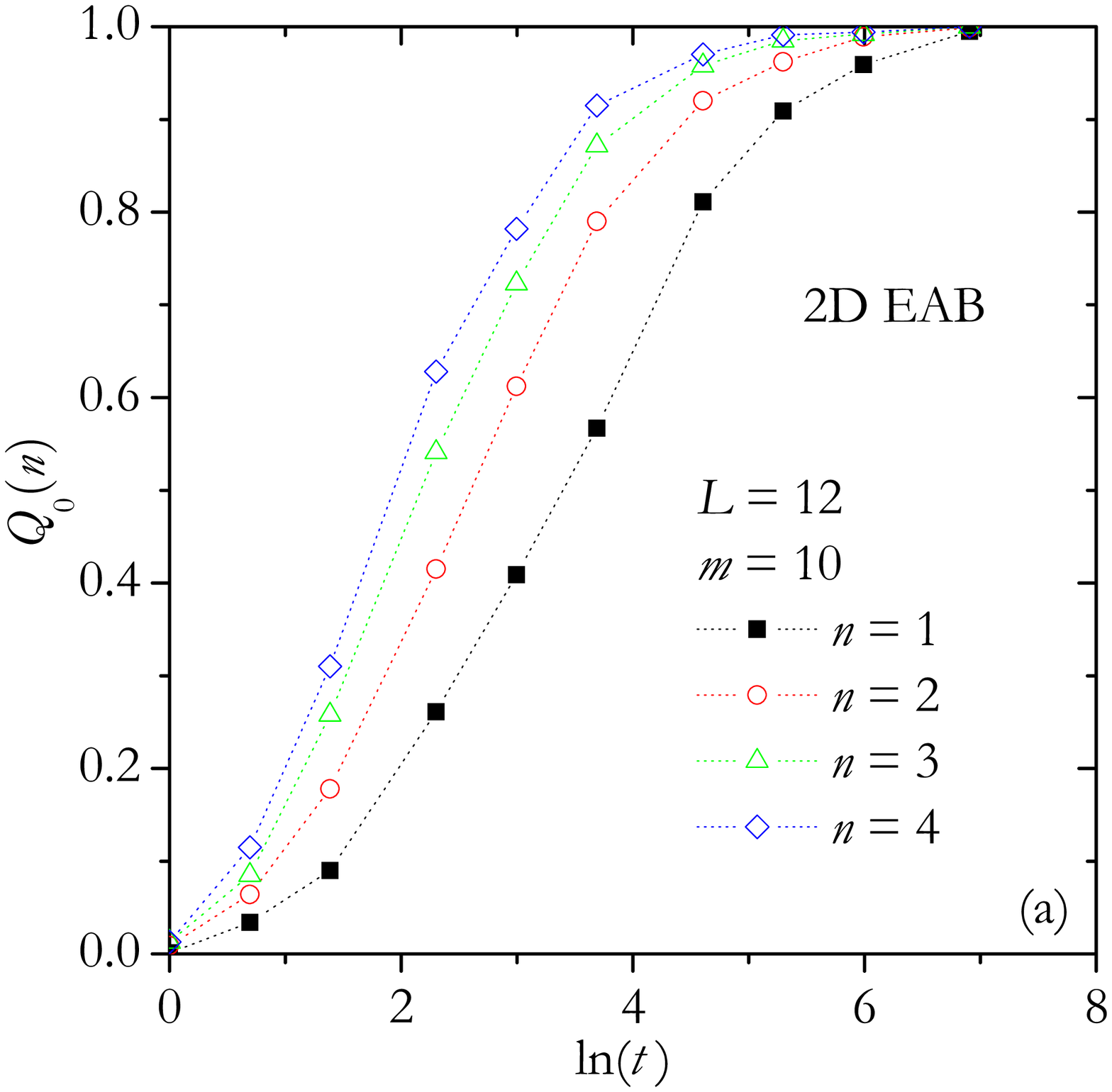}
\includegraphics[width=8cm,clip=true]{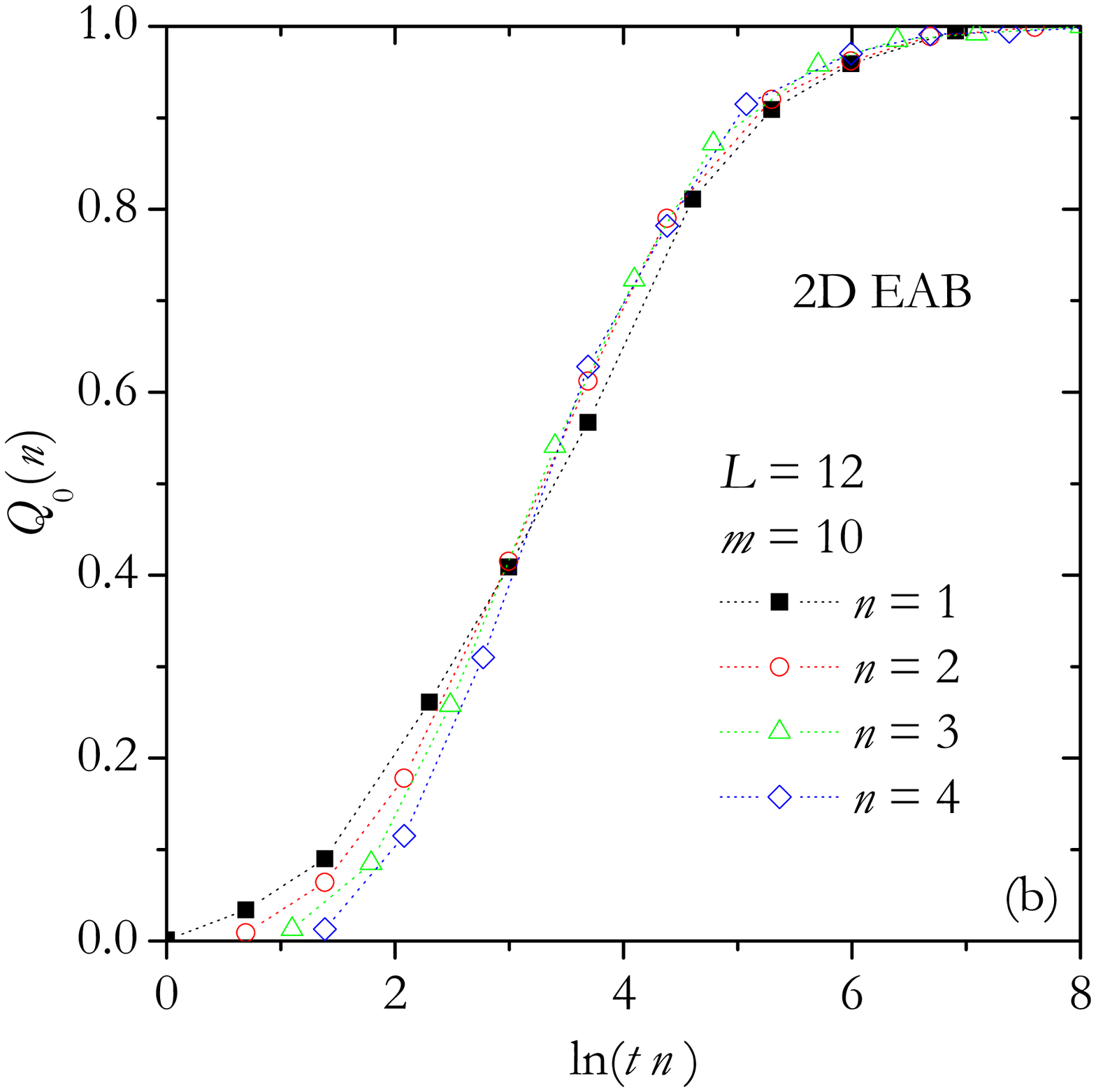}
\caption{\label{figure2}  $Q_0(n)$ for the 2D EAB model with
$L=12$, $m=10$ and different values of $n$ as indicated. (a)
$Q_0(n)$ vs $\ln(t)$, and (b) $Q_0(n)$ vs $\ln(t ~ n)$.}
\end{figure}

In this section we study the A algorithm for the EAB and EAG
models in 2D and 3D. The first issue we address is whether it is
better to use a large $t$ and one run for each sample, or several
runs but with a smaller $t$. The quantity to be studied for this
purpose is $\mathcal{Q}_0(n)$, which is the sample average of the
probability that a GS is found in at least one of the $n$
independent runs in the $j$-th sample, $\mathcal{Q}_{0,j}(n)$.
Figure~\ref{figure2} (a) shows the estimate of this quantity,
$Q_0(n)$, as a function of $\ln(t)$ for $n=1$, $2$, $3$ and $4$.
We use $N_{\mathrm{s}}=10^3$ samples of $L=12$ and $m=10$.  As
expected, the performance improves with increasing $n$. However,
when time is rescaled to $\ln(t ~ n)$ in order to compare the
performances at the same running time [see Eq.~(\ref{tsec})],
independently of the number of runs, the curves approximately
collapse [see Fig.~\ref{figure2} (b)]. The example shows that
increasing the number of PTSs, has approximately the same effect
as performing independent runs.  This is due to the fact that, if
an appropriate set of parameters are chosen, the fractions of
phase space explored by the PT algorithm are similar in both
cases. As this is a recurrent feature in all tested problems, in
the rest of this article we take $n=1$.

In the following, we first determine the range of temperatures
which is {\it globally optimal}. Although for each particular
problem (EAB or EAG in either 2D or 3D) is possible to determine a
different optimal set, for simplicity we choose to fix the range
of temperatures. Then, we focus on the number of replicas and the
time $t$. Finally, the analysis for small size allows us to
predict the optimal values of these parameters for larger lattice
sizes.

\subsection{2D EA models}

We begin by discussing the criteria for selecting a suitable range
of temperature for each studied system. As mentioned above, it is
important that the highest temperature $T_1$ is set in the
high-temperature phase: $T_1>T_{\mathrm{c}}$, where
$T_{\mathrm{c}}$ is the critical temperature. Although for the 2D
EA models $T_{\mathrm{c}}=0$, below $T\approx 1.3$ the dynamics
becomes slow \cite{Binder1986} and the system has very long
relaxation times. Then, it is reasonable to expect that the
optimal $T_1$ should be $T_1>1.3$. On the other hand, as the
algorithm is designed to reach the GS, the lowest temperature
$T_m$ should be very low (but not zero because the Metropolis rule
is not efficient in that case).

\begin{figure}[t]
\includegraphics[width=8cm,clip=true]{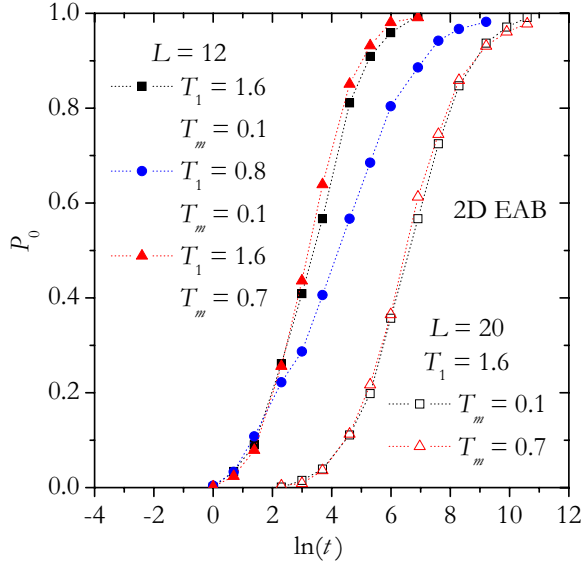}
\caption{\label{figure3}  $P_0$ as function of $\ln (t)$ for the
2D EAB model with $m=10$ and for sizes $L=12$ and $L=20$.  The
curves correspond to different ranges of temperatures as
indicated. }
\end{figure}

As in the previous section, we run the algorithm with several
different temperature ranges, on $N_{\mathrm{s}}=10^3$ different
samples per size (in the following, this number of samples is used
for all 2D calculations). $P_0$ for the EAB model is shown in
Fig.~\ref{figure3}. We find that, if we fix the lowest temperature
at $T_m=0.1$, for the case $L=12$ the highest temperature should
not be lower than $T_1=1.3$. On the other hand, if we choose the
highest temperature at $T_1=1.6$, the lowest temperature should
not be higher than $T_m=0.7$. In addition, as seen for $L=20$,
both ranges $T_1=1.6$ to $T_m=0.1$ and $T_1=1.6$ to $T_m=0.7$,
give similar results. The conclusion is that the performance of
the algorithm depends only weakly on the range of temperatures
chosen under the condition that, a) the largest temperature is
outside the region of slow dynamics ($T_1>1.3$), and b) the lowest
temperature is not too high (say, $T_m<0.7$). This conclusion is
also valid for the EAG model. As the bounds mentioned are only
approximate, to be sure to meet that condition, in the remaining
of this subsection we use $T_1=1.6$ and $T_m=0.1$ for all the
simulations.

\begin{figure}[t]
\includegraphics[width=8cm,clip=true]{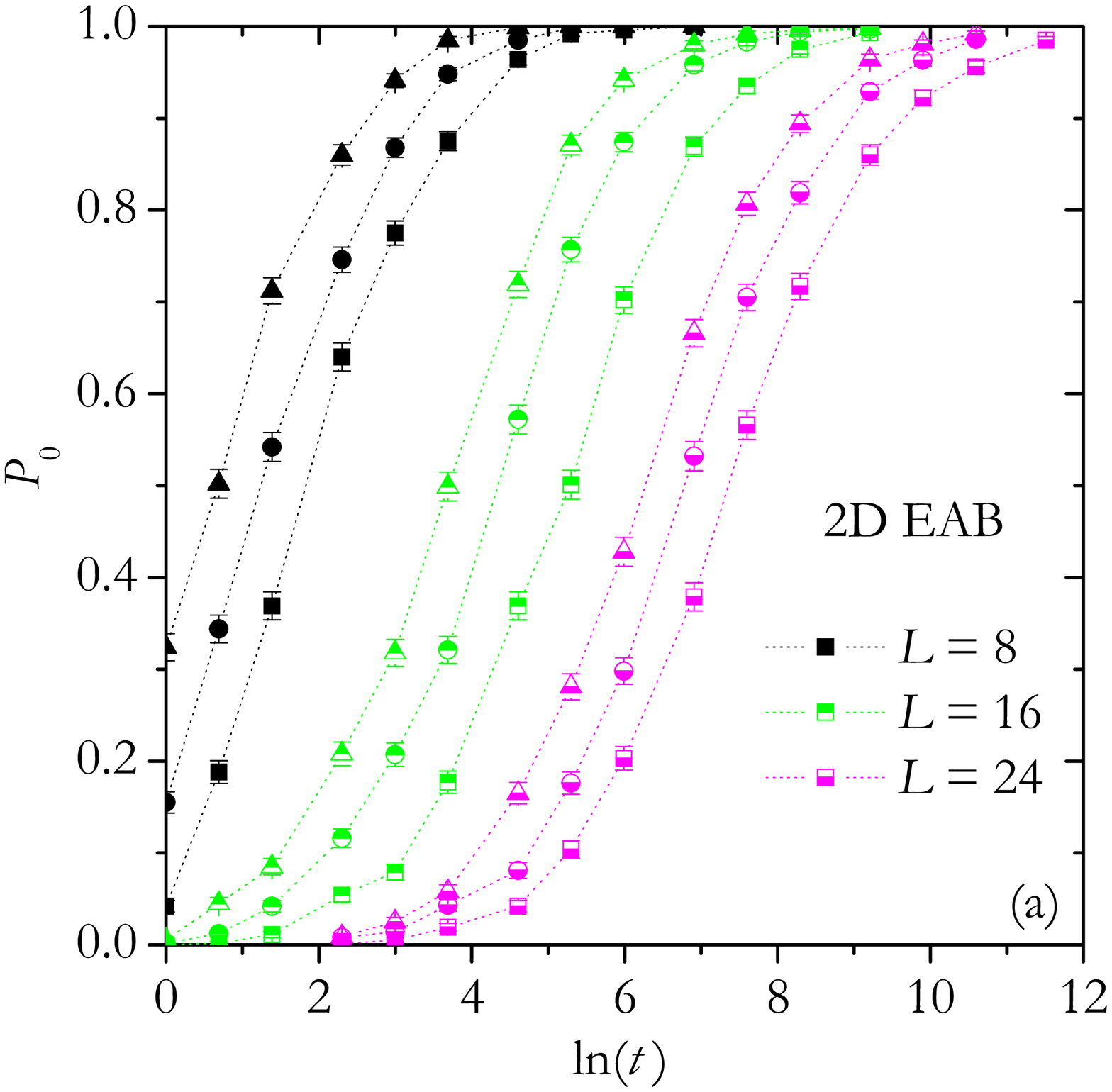}
\includegraphics[width=8cm,clip=true]{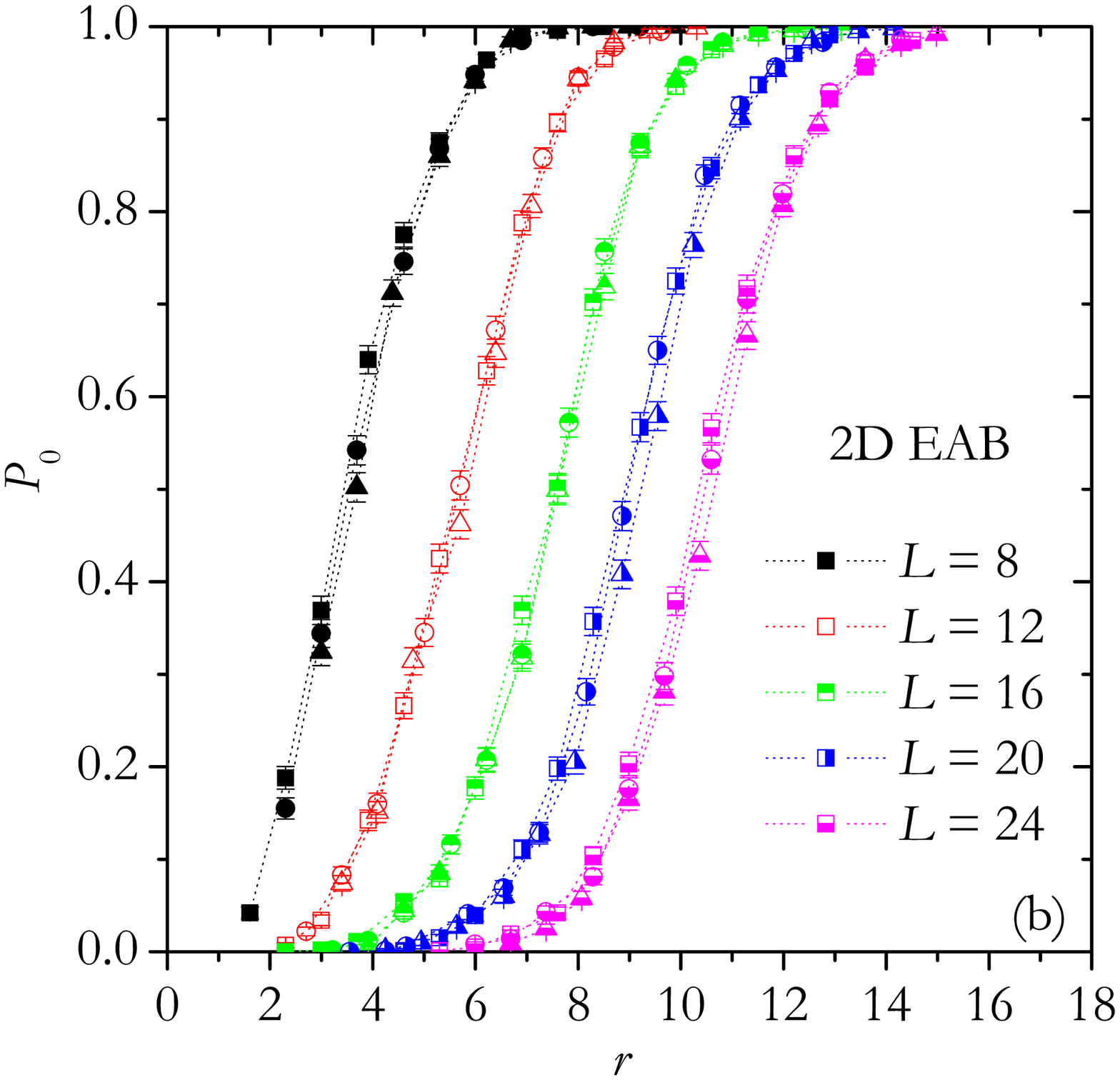}
\caption{\label{figure4}  $P_0$ as function of (a) $\ln(t)$ and
(b) $r$ for the 2D EAB model.  Squares, circles and triangles are
results for, respectively: $L=8$ with $m=5$, $10$ and $20$; $L=12$
with $m=5$, $15$ and $30$; $L=16$ with $m=10$, $25$ and $50$;
$L=20$ with $m=10$, $35$ and $70$; $L=24$ with $m=20$, $40$ and
$80$ (the values of $m$ are given from right to left). }
\end{figure}

We now discuss the dependence of probability $P_0$ on parameters
$L$, $m$ and $t$. Figure~\ref{figure4} (a) shows the curves of
probability for different lattice sizes and number of replicas. As
was to be expected, for fixed values of $t$, $P_0$ increases with
$m$. Thus, given that the running time is proportional to $t~m$
[see Eq.~(\ref{tsec})], is reasonable to draw the curves as
functions of
\begin{equation}
r \equiv \ln(t ~ m) \label{r},
\end{equation}
as in Fig.~\ref{figure4} (b).  As before, when we carry out
independent runs, the curves for different $m$ and fixed $L$
collapse to the same curve.  This behavior shows that $P_0$ is a
function of $r$ for fixed size \cite{Nota1}.

The collapse observed previously can be used to find a formula,
that allows us to estimate the number of PTSs that are necessary
to reach the GS with a {\it average} probability $P_0$. The
simplest function that we have found to provide a good fit for the
data has five parameters. In the following paragraphs we show,
using heuristic arguments, how such a formula is obtained.

\begin{figure}[t]
\includegraphics[width=8cm,clip=true]{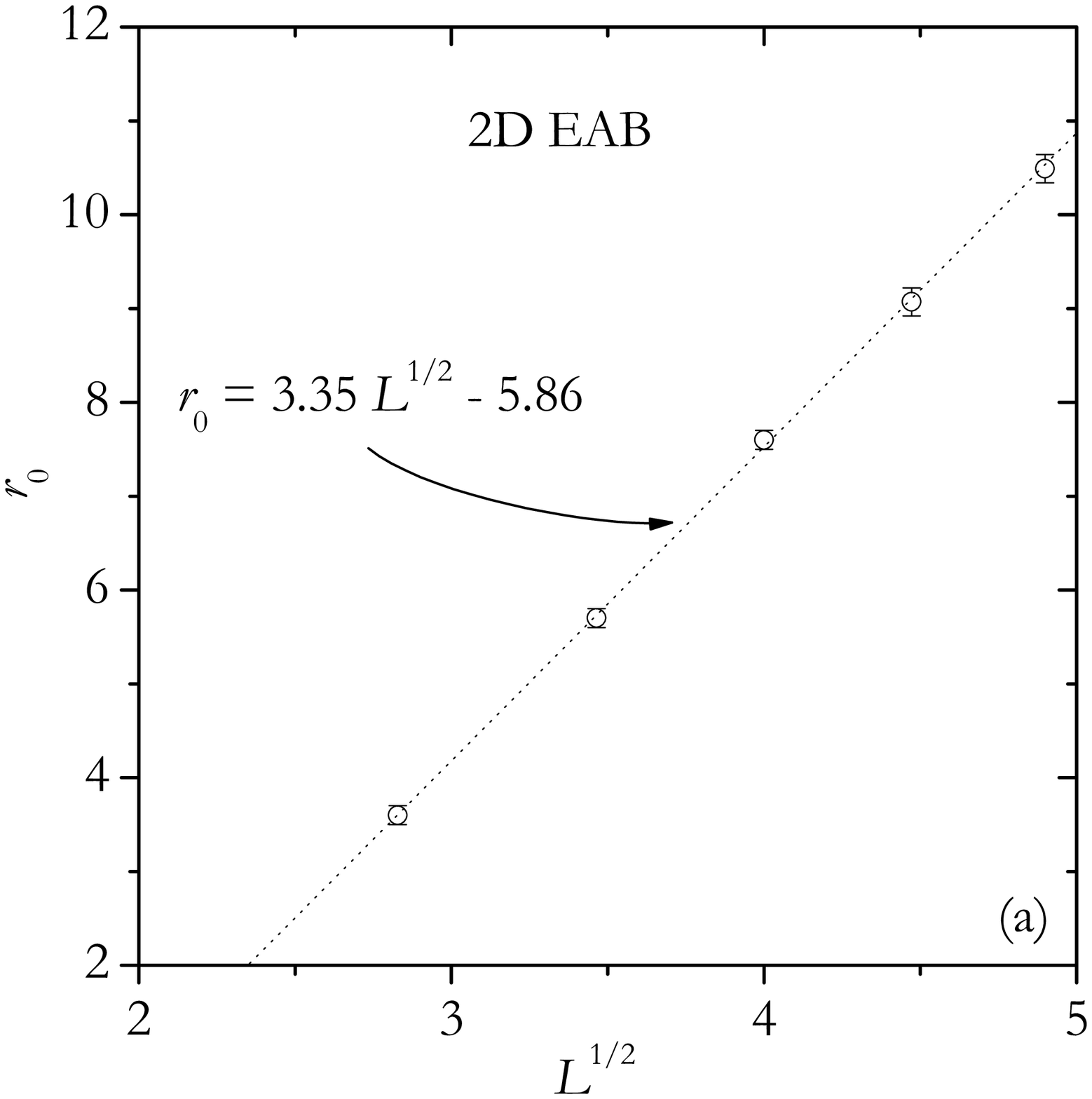}
\includegraphics[width=8cm,clip=true]{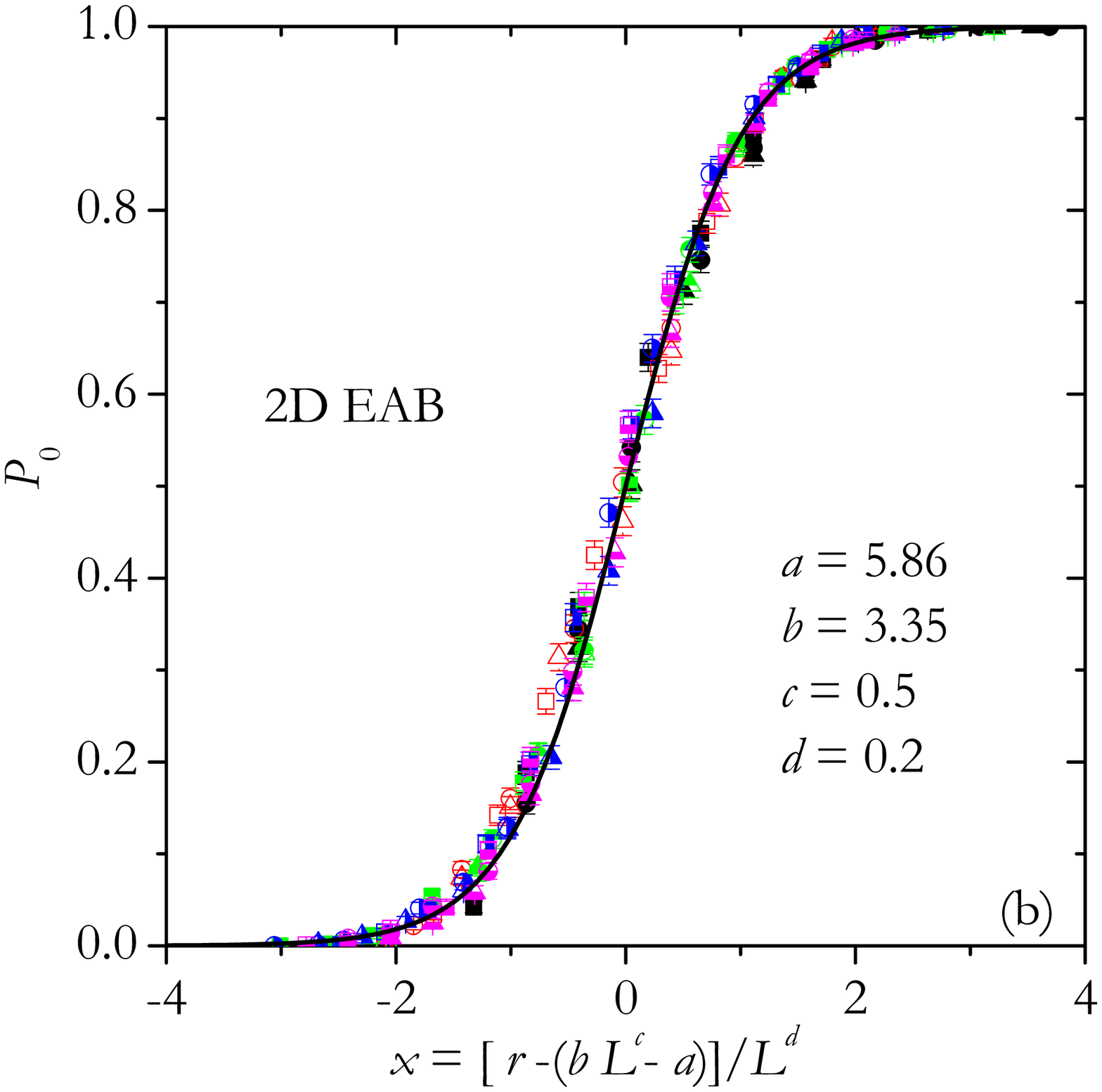}
\caption{\label{figure5}  (a) $r_0$ vs $L^{1/2}$ for the 2D EAB
model.  (b) Data collapsing for all curves in Fig.~\ref{figure4}
(b).  The full line is the function defined in Eq.~(\ref{f}) with
$q=2$.}
\end{figure}

We now turn to the $L$ dependence of the curves shown in
Fig.~\ref{figure4} (b), with the hope that they too can be
collapsed onto a single curve. First, note that the inflection
point is located approximately at $P_0=0.5$. Figure~\ref{figure5}
(a) shows that $r_0$, the value of $r$ at $P_0=0.5$, can be
approximated by a linear function of $L^{1/2}$ (a poor fit is
obtained if a linear function of $L$ is used).  By fitting these
points with the function
\begin{equation}
r_0 =bL^c-a,
 \label{fit1}
\end{equation}
we obtain $b=3.35(7)$ and $a=5.86(28)$ ($c$ is fixed to $c=1/2$,
but it is left as a variable in the equation because its value is
different in 3D).  We rescale the abscissa axis of
Fig.~\ref{figure4} (b), using the variable
\begin{equation}
x=[r - (bL^c-a)]/L^d.
 \label{x}
\end{equation}
The denominator has been introduced to compensate for the fact
that the slope at $r_0$ changes slightly with $L$.
Figure~\ref{figure5} (b) shows that, for $d=0.2$, this rescaling
indeed collapses all the curves of figure~\ref{figure4} (a) onto a
single curve.

Furthermore, we find that all data points of Fig.~\ref{figure5}
(b) can be fitted by the function
\begin{equation}
f(x)=\frac{\exp(qx)}{1+\exp(qx)},
 \label{f}
\end{equation}
where $q=2$ [see figure~\ref{figure5} (b)]. By combining
Eqs.~(\ref{x}) and (\ref{f}), we can obtain a simple expression to
predict the number of PTSs necessary to reach the GS with a given
value of $P_0$,
\begin{equation}
t= \frac{1}{m}
\left(\frac{P_0}{1-P_0}\right)^{\frac{L^d}{q}}\exp(bL^c-a).
 \label{tm}
\end{equation}
Even though the rescaling proposed is not rigorous and the
function Eq.~(\ref{f}) is chosen arbitrarily, the expression
Eq.~(\ref{tm}) will allow us to estimate the value of $t$ for each
lattice size $L$ and for a given average probability $P_0$ with
great accuracy (see Subsec.~\ref{sec3C}). Notice that an
exponential dependence on size has also already been reported in
Ref.~\cite{Katzgraber2003b}, for the time that a PT algorithm
needs to find the GS of a 1D spin glass model with power law
interactions, in the case where the power law parameter is $\sigma
\approx 2.5$, which can be considered effectively as a short range
interaction.

We stress that the purpose of Eq.~(\ref{tm}) is not to provide an
accurate model of the PT algorithm, but only to give a simple
estimate of the number of PTSs needed to achieve a certain {\it
average} probability of finding a GS. In fact, the actual $P_0$
obtained using this estimate can vary strongly from sample to
sample (see Sec.~\ref{sec4}).

\begin{figure}[t]
\includegraphics[width=8cm,clip=true]{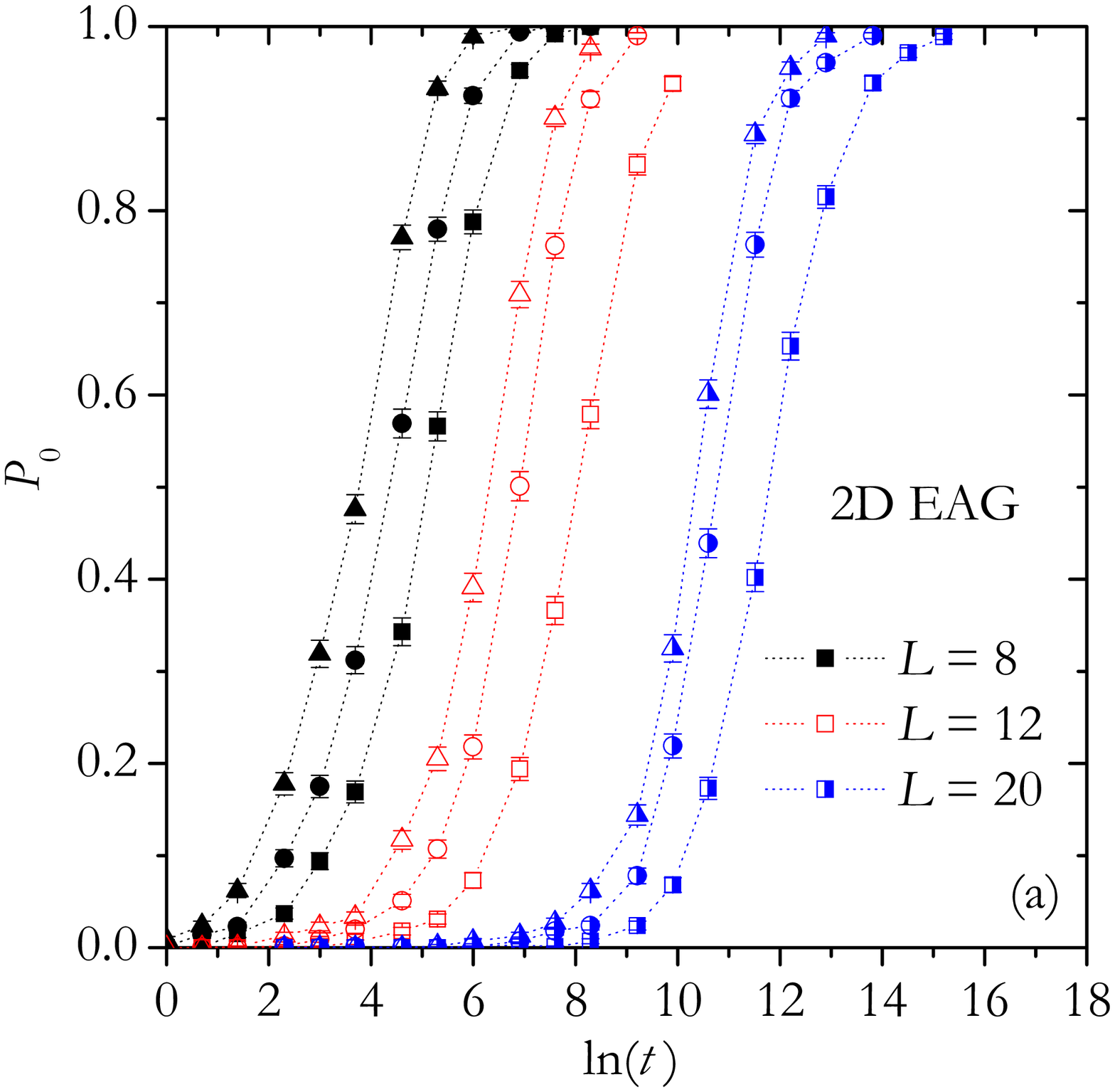}
\includegraphics[width=8cm,clip=true]{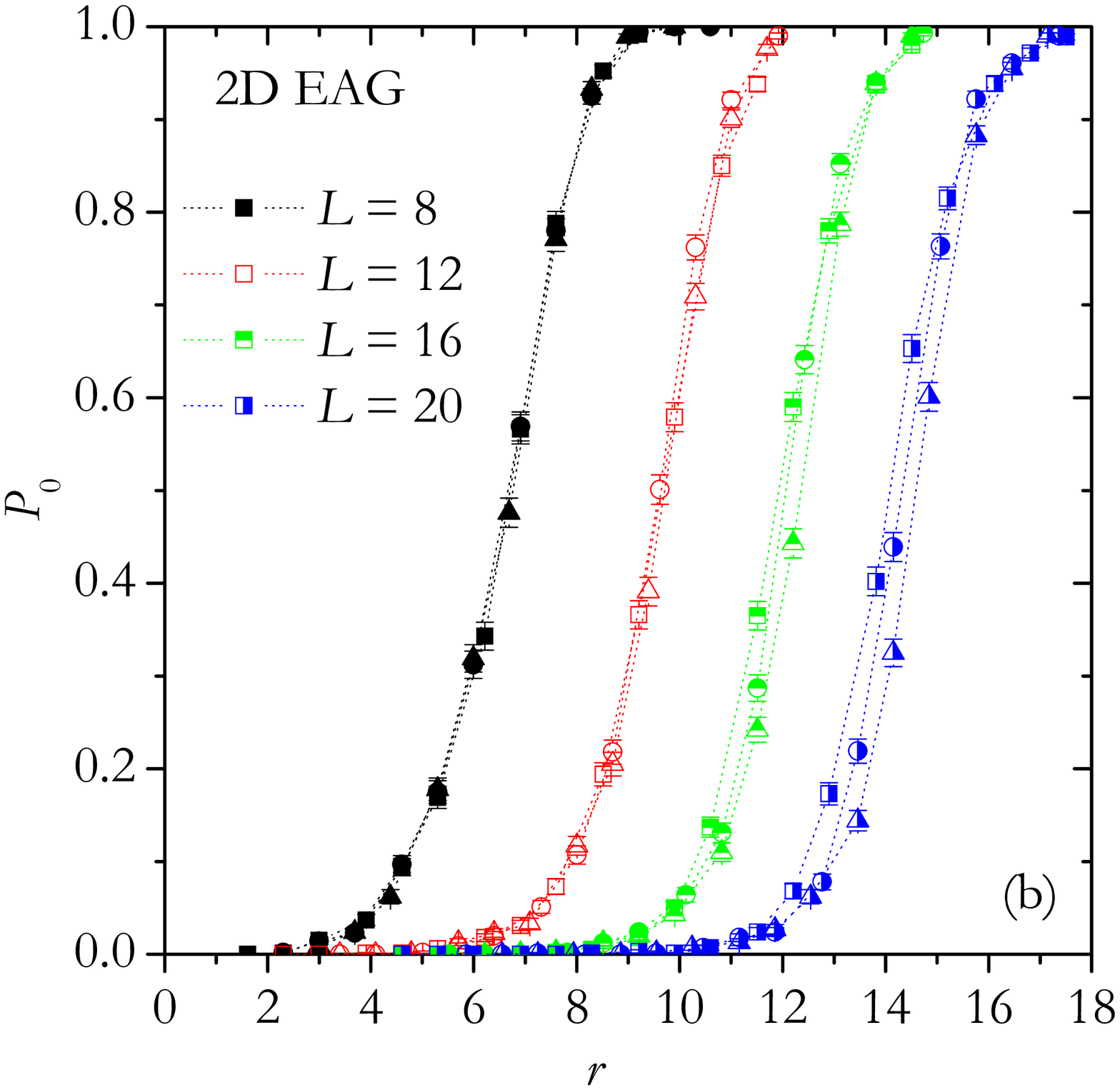}
\caption{\label{figure6}  $P_0$ as function of (a) $\ln(t)$ and
(b) $r$ for the 2D EAG model. Squares, circles and triangles are
results for, respectively: $L=8$ with $m=5$, $10$ and $20$; $L=12$
with $m=5$, $15$ and $30$; $L=16$ with $m=10$, $25$ and $50$;
$L=20$ with $m=10$, $35$ and $70$ (the values of $m$ are given
from right to left). }
\end{figure}

\begin{figure}[t]
\includegraphics[width=8cm,clip=true]{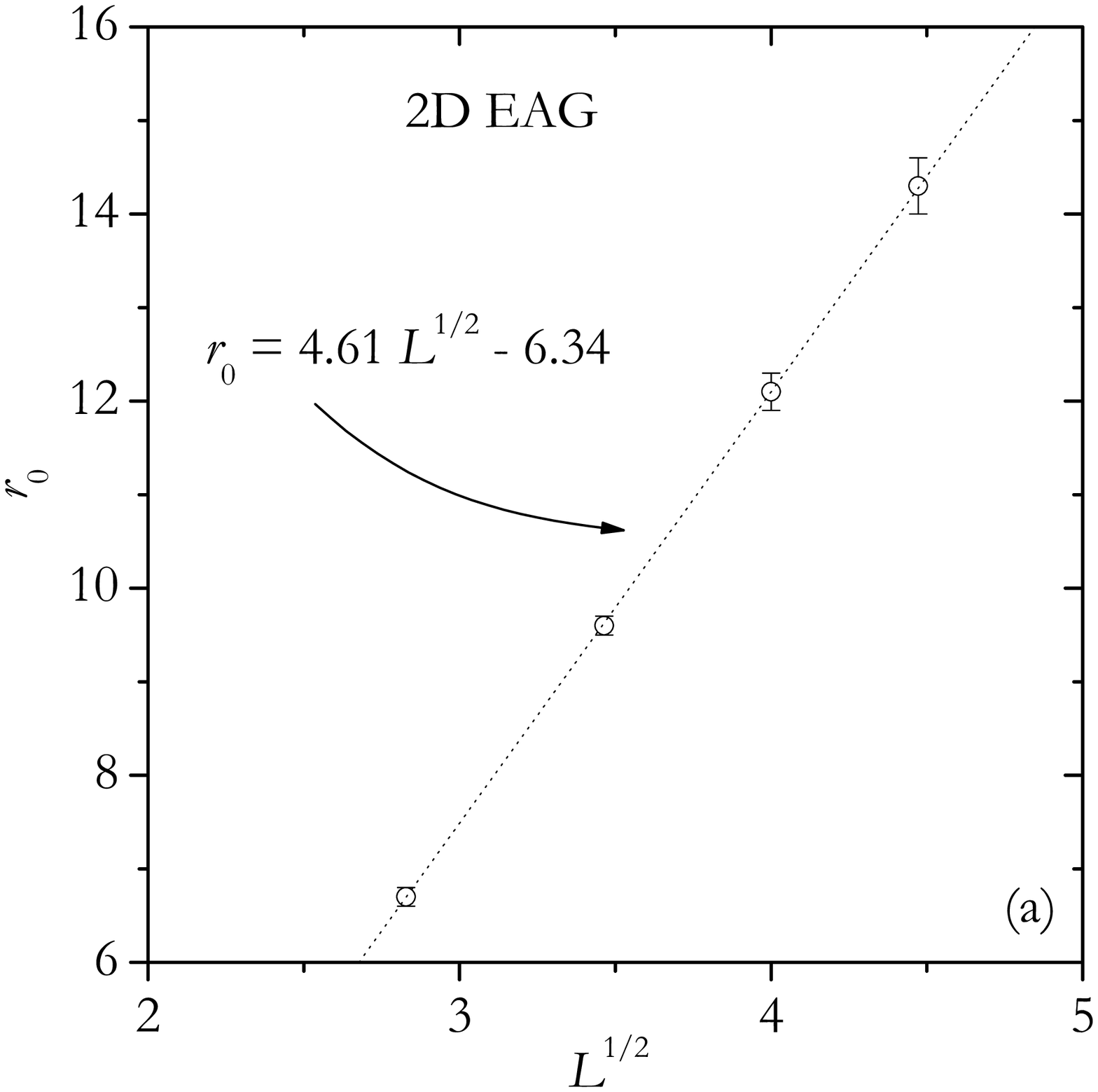}
\includegraphics[width=8cm,clip=true]{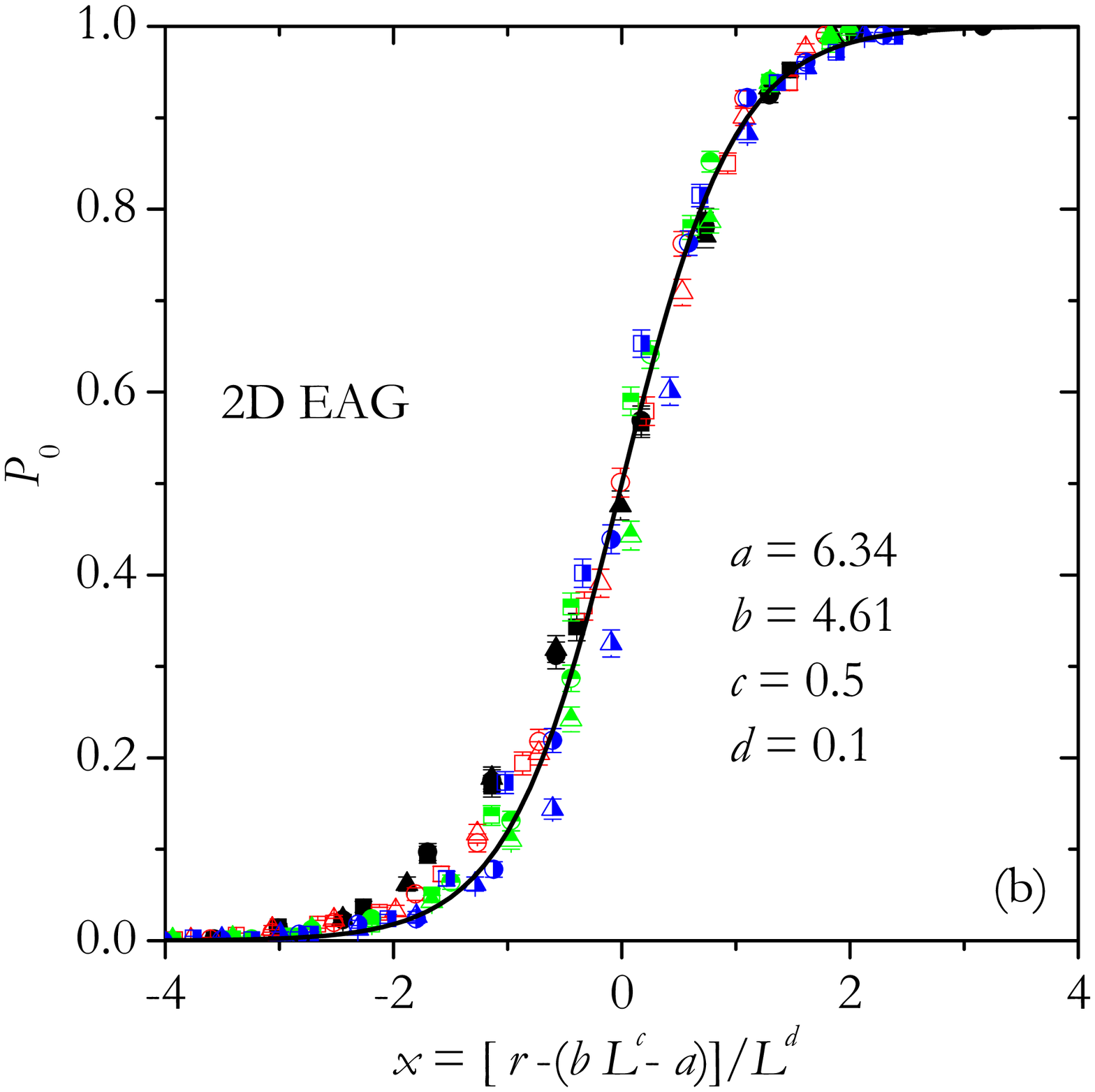}
\caption{\label{figure7}  (a) $r_0$ vs $L^{1/2}$ for the 2D EAG
model.  (b) Data collapsing for all curves in Fig.~\ref{figure6}
(b). The full line is the function defined in Eq.~(\ref{f}) with
$q=2$.}
\end{figure}

Now, we repeat the previous analysis for the EAG model.
Figures~\ref{figure6} (a) and (b) show the probability $P_0$ for
different lattice sizes and number of replicas as functions of
$\ln(t)$ and $r$, respectively. Even though the collapse in this
case is not as good as for the EAB model, it is still very good in
the region of high $P_0$ ($P_0 \gtrsim 0.9)$. Figure~\ref{figure7}
(a) shows $r_0$ as function of $L^{1/2}$. As before, we fit these
points with Eq.~(\ref{fit1}) and we obtain $b=4.61(14)$ and
$a=6.34(46)$ ($c=1/2$). A second collapsing of the data is shown
in Fig.~\ref{figure7} (b), again for $d=0.2$. This collapse is
well fitted by the function in Eq.~(\ref{f}) with $q=2$ [see
figure~\ref{figure7} (b)].

One difference with the EAB case is that the value of the
parameter $b$ is larger, which implies that, for example, for
$L=20$, $m=10$ and $P_0 \approx 0.99$, the required number of PTSs
for the EAG model is two orders of magnitude larger than for the
EAB model. This is also a common feature of other heuristics as
genetic algorithm, where GSs are harder to obtain for Gaussian
than for bimodal bond distributions. The reason is that the GS of
the EAG model is unique, while it is degenerate for the EAB model,
making it easier to find because any one of them is sufficient
\cite{Pal1996b}. In fact, the ground-state entropy per spin for
the EAB model with $L=20$ is approximately $S=0.0818$
\cite{Roma2004}, what implies that the number of GSs is ~$\sim 1.6
\times 10^{14}$.

\subsection{3D EA models}

In the previous analysis for 2D lattices we have checked the GS
energies obtained by PT, comparing them with the exact ones
calculated with branch-and-cut algorithm \cite{DeSimone1995}.  But
finding the GS of a 3D system is a much more difficult task, and
exact algorithms available to us can only be used for small
lattices (up to $L=6$). Therefore, a different strategy must be
used to ensure that the configurations found by our algorithm
really correspond to true GSs. The method we have used is as
follows. First, we run the algorithm for a certain time $t$, for
each sample of a given set. Then, the algorithm is run anew but
now duplicating the time $t$. This is repeated many times, and
thus a series of configurations, with their energies, are stored.
We stop this process when, for each sample, in two successive runs
the same minimum energy is obtained. Note that the process
continues in all samples, while at least the minimum energy for
one sample changes in two successive runs. Comparing the series of
energies, we can separate them into "easy" and "hard" samples (the
GS energy of an easy sample is obtained in a few steps of the
previous process, while for a hard sample many more steps are
necessary). In this stage, we assume that the true GS has been
found for all easy samples. On the other hand, for hard samples
the previous process is continued and it is assumed that the true
GS has been found when in three successive runs the same energy is
obtained.  We used these energies to calculate the probability
$P_0$ for all 3D samples studied in this subsection. Note that the
number of PTSs used for reaching the GS is not necessarily
optimal, i.e. it is not impossible that the GS can be found in
shorter times.

\begin{figure}[t]
\includegraphics[width=14cm,clip=true]{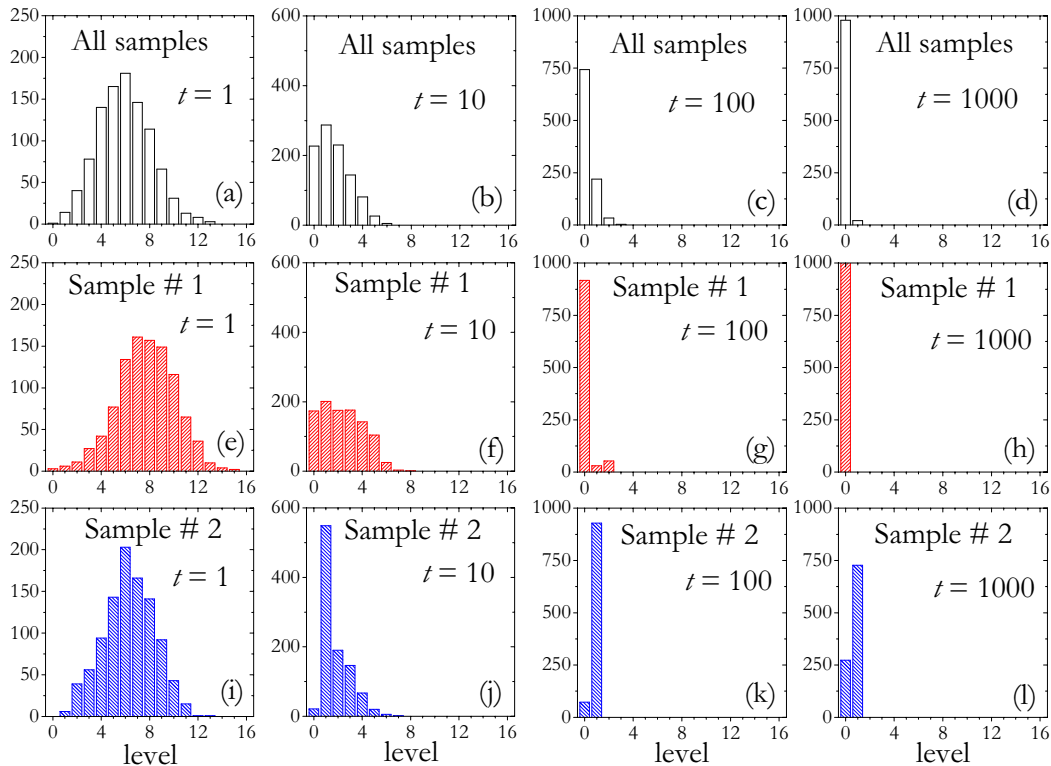}
\caption{\label{figure8}  Energy levels found with the PT
algorithm for the 3D EAB model with $L=6$, $m=20$ and times $t=1$,
$10$, $10^2$ and $10^3$. Histograms for (a)-(d)
$N_{\mathrm{s}}=10^3$ different samples ($1$ run per sample),
(e)-(h) $10^3$ runs of an easy sample ($\#1$) and (i)-(l) $10^3$
runs of a hard sample ($\#2$).}
\end{figure}

\begin{figure}[t]
\includegraphics[width=8cm,clip=true]{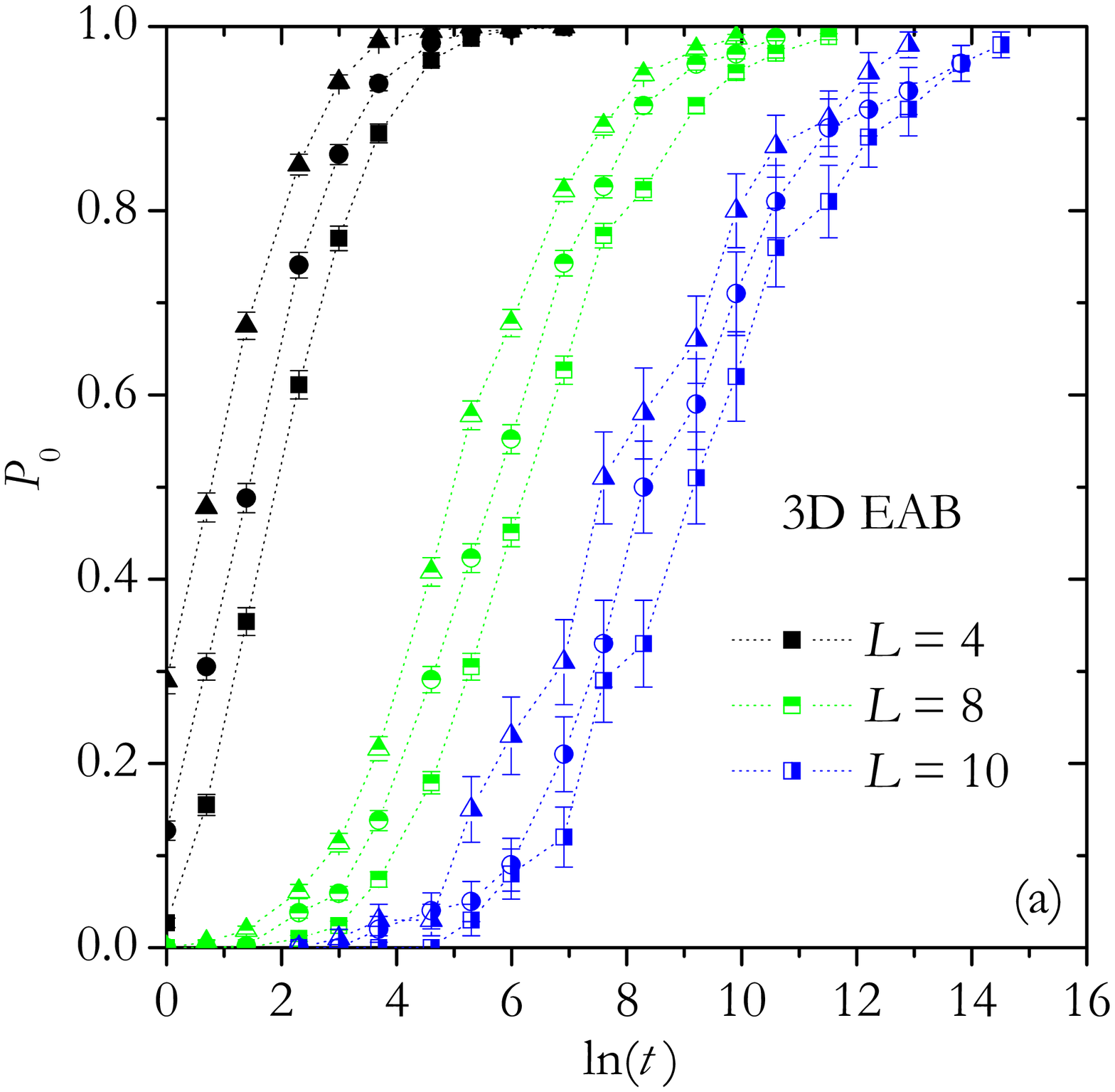}
\includegraphics[width=8cm,clip=true]{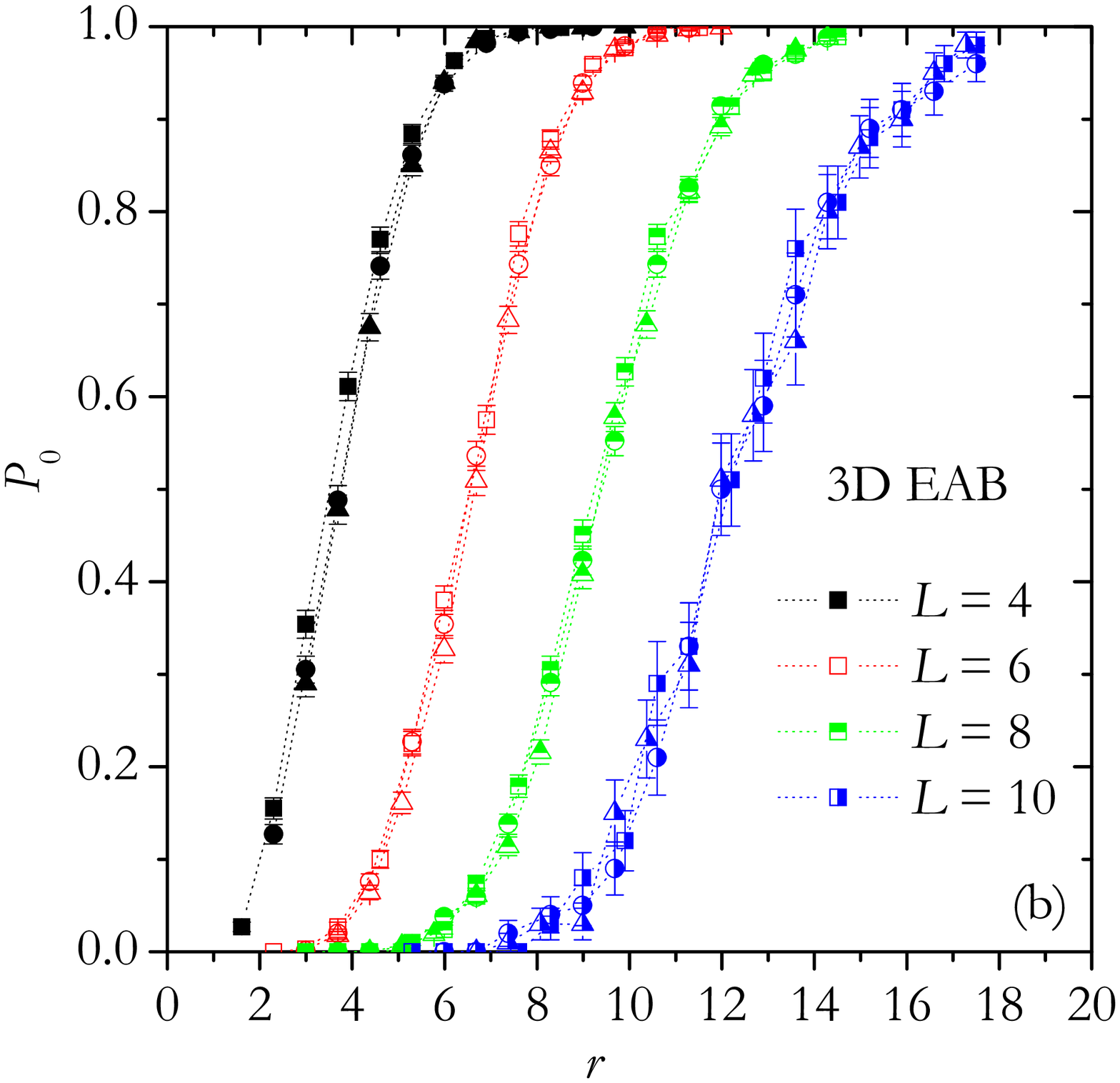}
\caption{\label{figure9}   $P_0$ as function of (a) $\ln(t)$ and
(b) $r$ for the 3D EAB model. Squares, circles and triangles are
results for, respectively: $L=4$ with $m=5$, $10$ and $20$; $L=6$
with $m=10$, $20$ and $40$; $L=8$ with $m=20$, $40$ and $80$;
$L=10$ with $m=20$, $40$ and $80$ (the values of $m$ are given
from right to left). }
\end{figure}

\begin{figure}[t]
\includegraphics[width=8cm,clip=true]{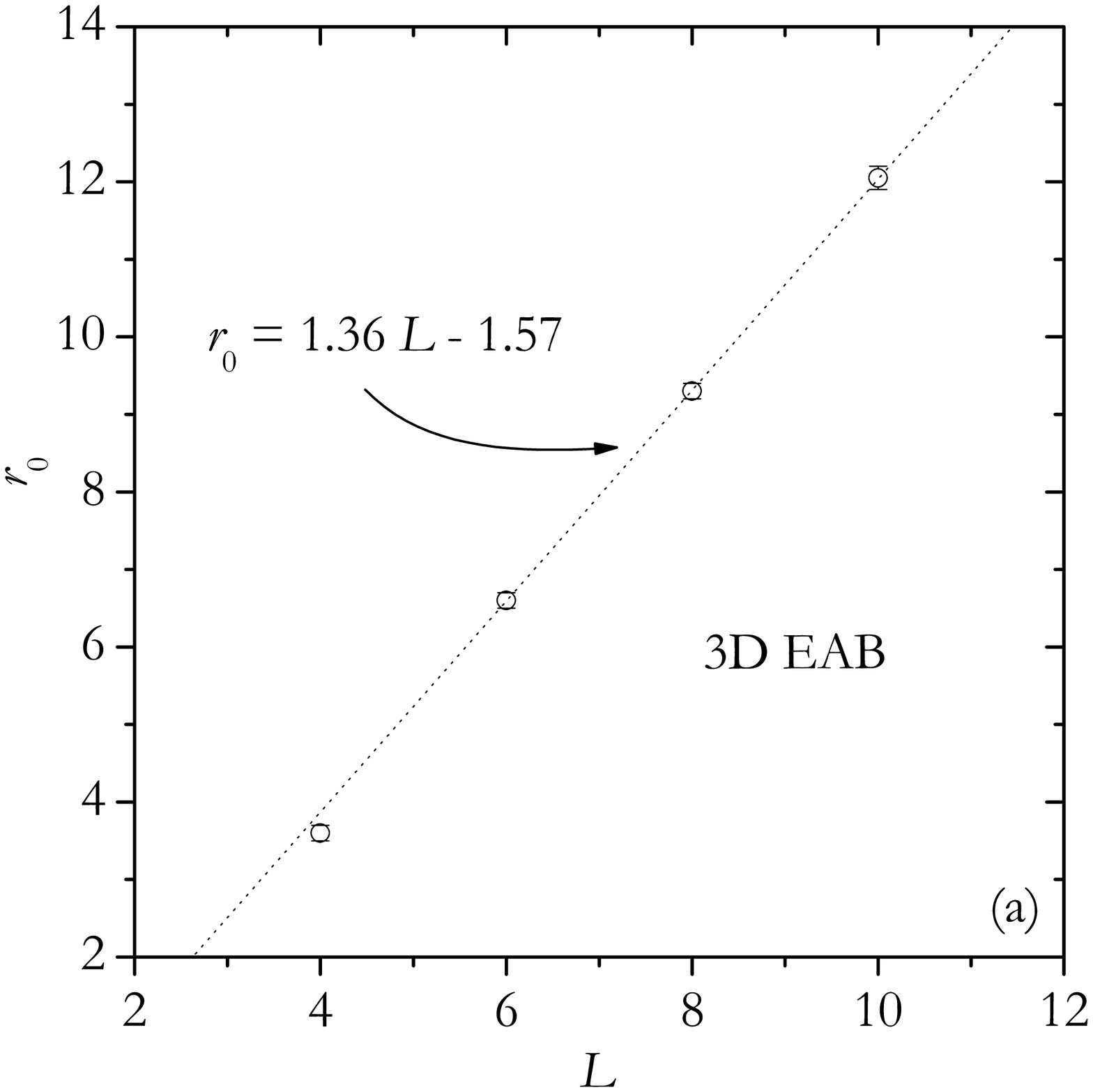}
\includegraphics[width=8cm,clip=true]{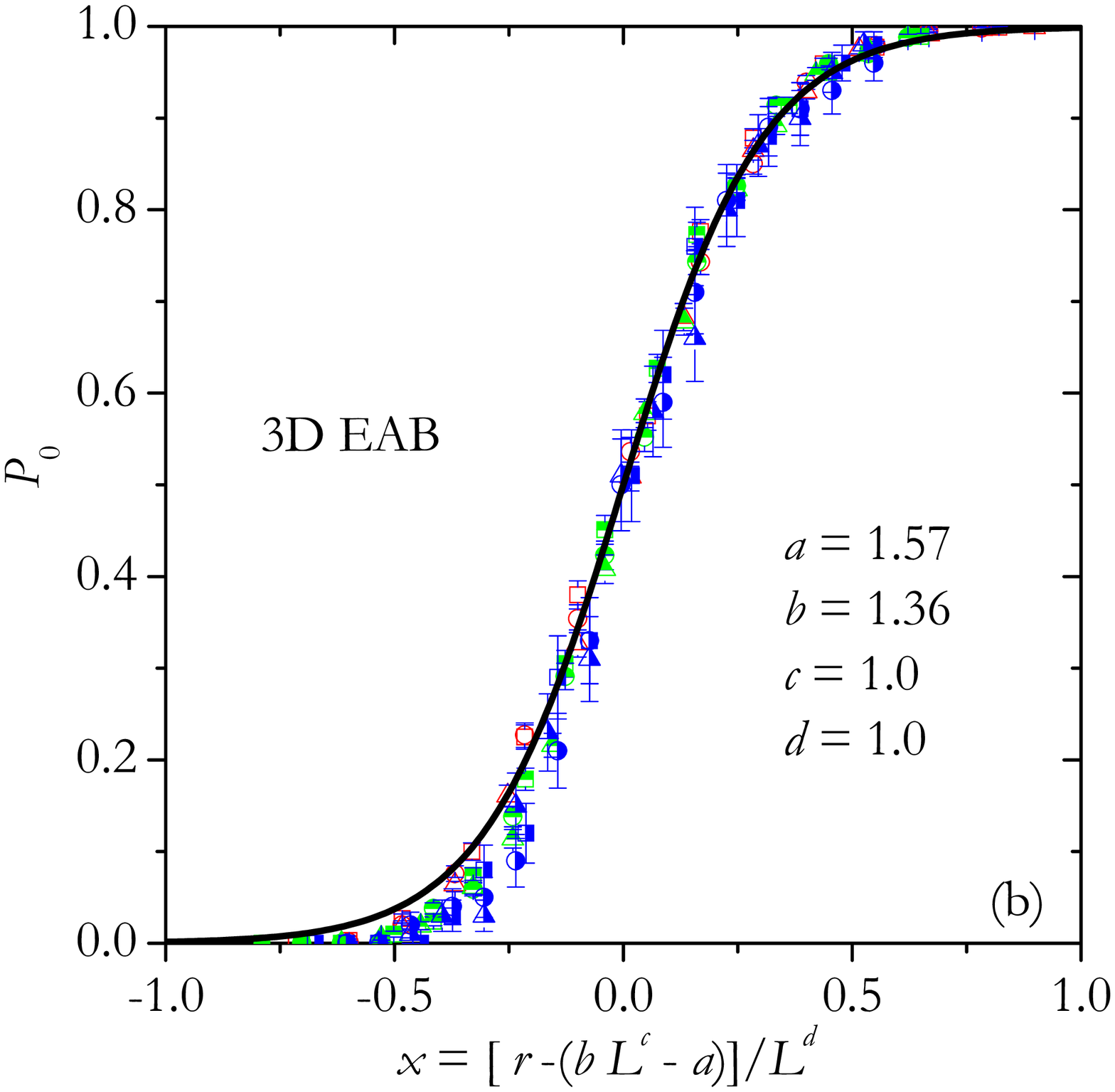}
\caption{\label{figure10}  (a) $r_0$ vs $L$ for the 3D EAB model.
(b) Data collapsing for all curves in Fig.~\ref{figure9} (b). The
full line is the function defined in Eq.~(\ref{f}) with $q=6.5$.}
\end{figure}

In Fig.~\ref{figure8} we can see the difference between the
histogram (number of samples) of the energy levels obtained with
the PT algorithm for many samples, and the ones obtained by
performing many independent runs of the algorithm on, an easy and
a hard sample. Figures~\ref{figure8} (a)-(d) show the histograms
for $N_{\mathrm{s}}=10^3$ samples of the 3D EAB model with $L=6$,
$m=20$ and four different numbers of PTSs: $t=1$, $10$, $10^2$ and
$10^3$ (one run for each sample). As it can be observed, for short
times such as $t=1$ and $t=10$, the histogram is broad and the
maximum is not located in the ground level. For long $t$, the
shape of the histogram changes and a peak arises at the ground
level. In fact, as $t=10^3$ is used, the GS energy is found in 979
samples (the remaining 21 are located in the first excited level).

A similar behavior is observed for the easy sample ($\#1$),
Fig.~\ref{figure8} (e)-(h).  In this case, instead of many
samples, one sample and $n=10^3$ independent run are used. For
$t=10^3$, the PT algorithm always finds a true GS. On the other
hand, as it is shown in Fig.~\ref{figure8} (i)-(l), a different
behavior is observed for the hard sample ($\#2$). For all $t$, the
peak is not located at the ground level. Thus, for $t=10^3$ the
true GS is found in only $273$ of the runs. This example shows
that the properties of hard samples are not reflected in the
global behavior [Figs.~\ref{figure8} (a)-(d)] and justifies our
previous protocol to obtain true GS.

In order to study the influence of the temperature range on the
performance of our algorithm, it is important to bear in mind that
for 3D $T_{\mathrm{c}}>0$.  For practical purposes, we can
consider that $T_{\mathrm{c}} \approx 1.12$ for the EAB model and
$T_{\mathrm{c}} \approx 0.95$ for the EAG model
\cite{Katzgraber2006}. After a similar analysis to the one carried
out above for 2D models, we conclude that $T_1=1.6$ and $T_m=0.1$
are the adequate limits for the 3D case and they will be used
throughout the section.

\begin{figure}[t]
\includegraphics[width=8cm,clip=true]{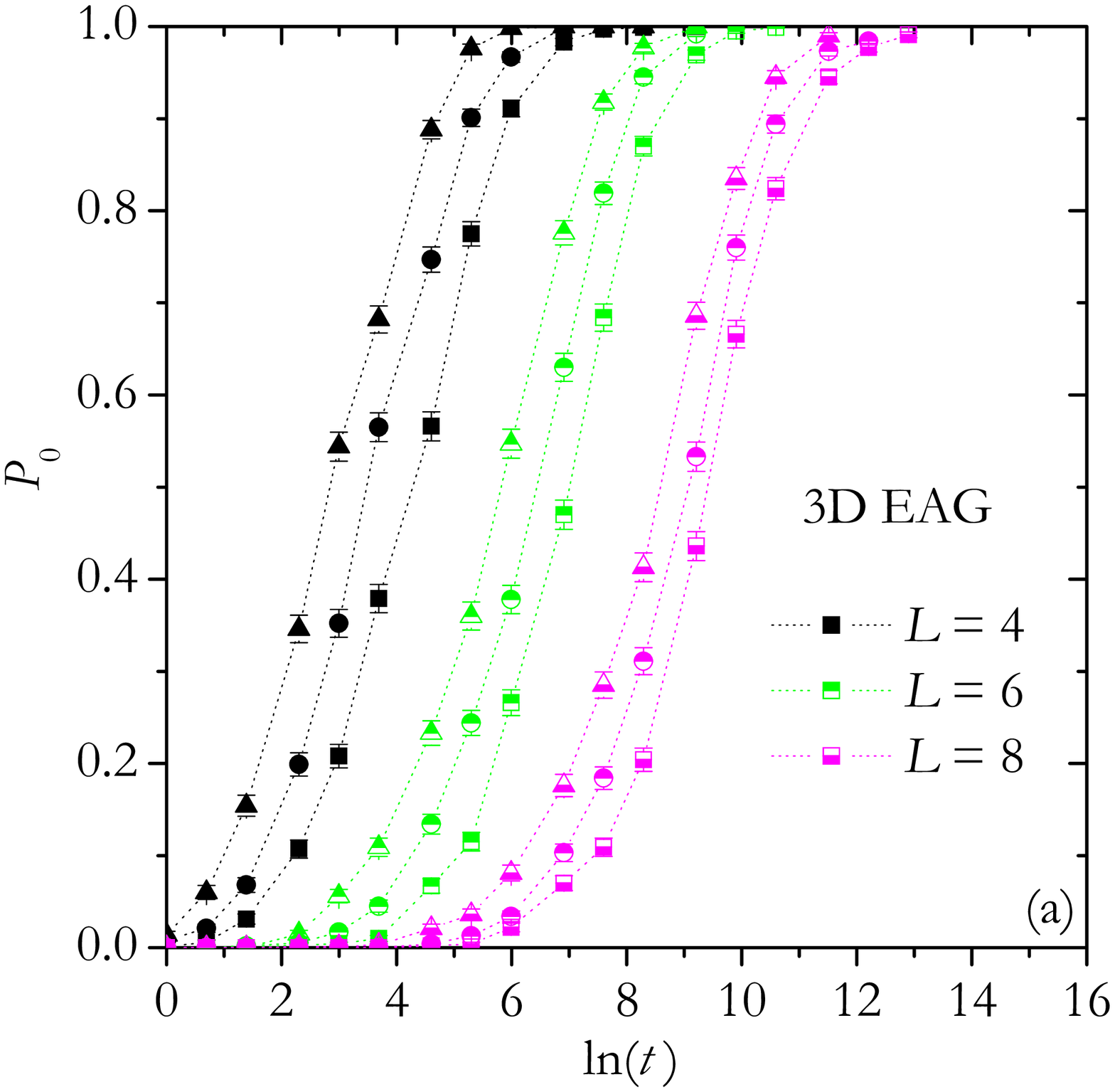}
\includegraphics[width=8cm,clip=true]{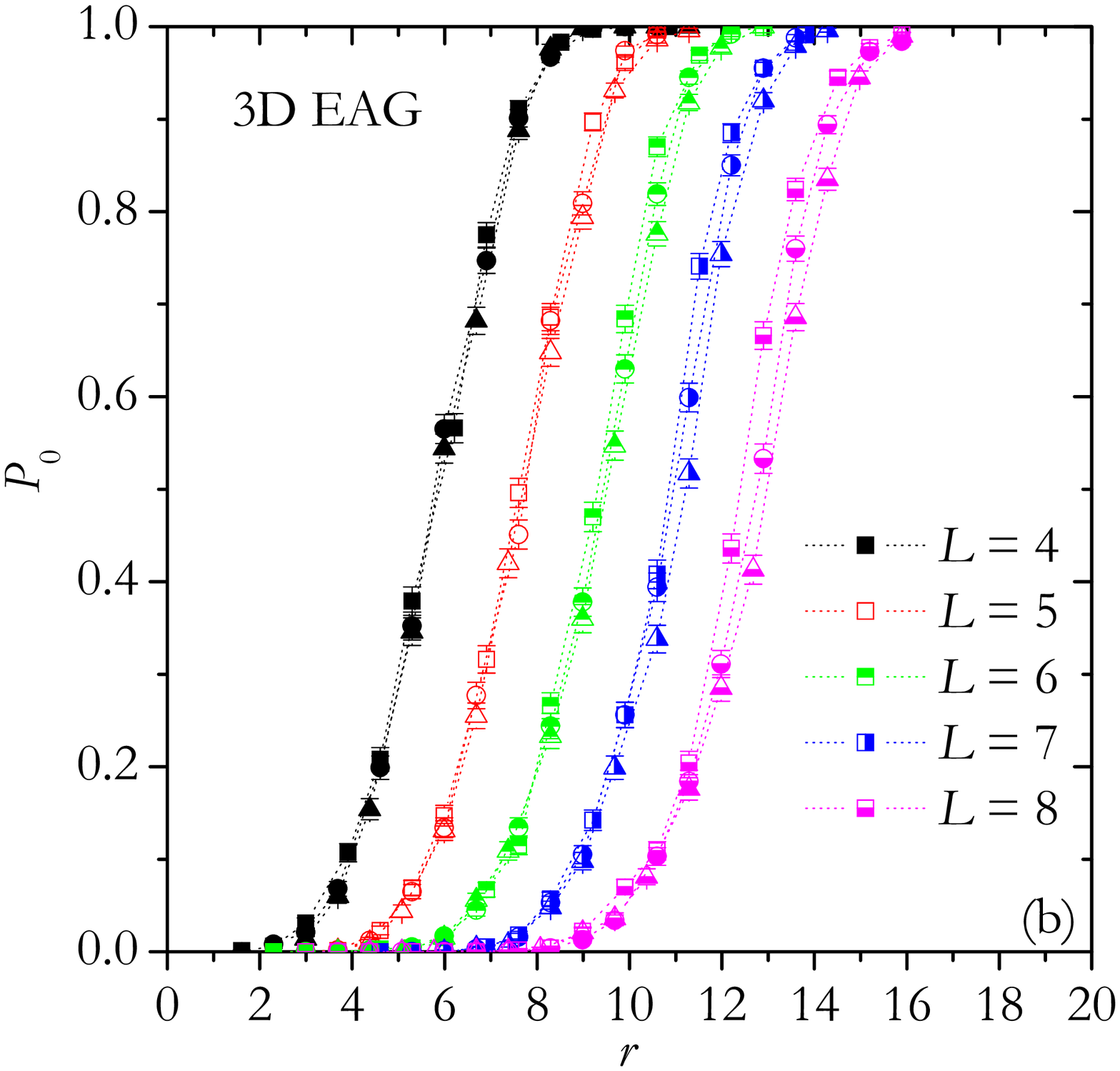}
\caption{\label{figure11}   $P_0$ as function of (a) $\ln(t)$ and
(b) $r$ for the 3D EAG model. Squares, circles and triangles are
results for, respectively: $L=4$ with $m=5$, $10$ and $20$; $L=5$
with $m=10$, $20$ and $40$; $L=6$ with $m=10$, $20$ and $40$;
$L=7$ with $m=10$, $20$ and $40$; $L=8$ with $m=20$, $40$ and $80$
(the values of $m$ are given from right to left). }
\end{figure}
\begin{figure}[t]
\includegraphics[width=8cm,clip=true]{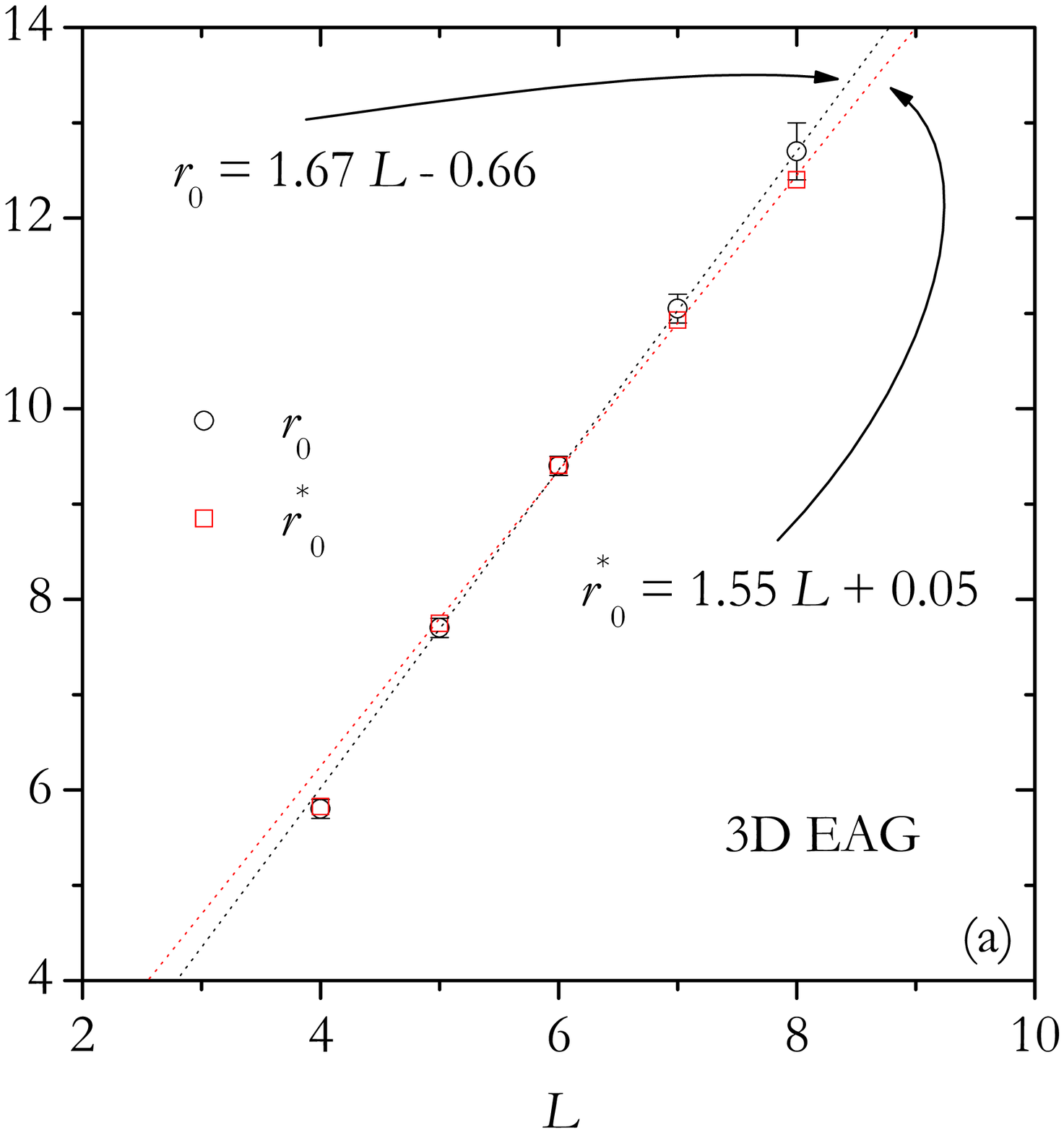}
\includegraphics[width=8cm,clip=true]{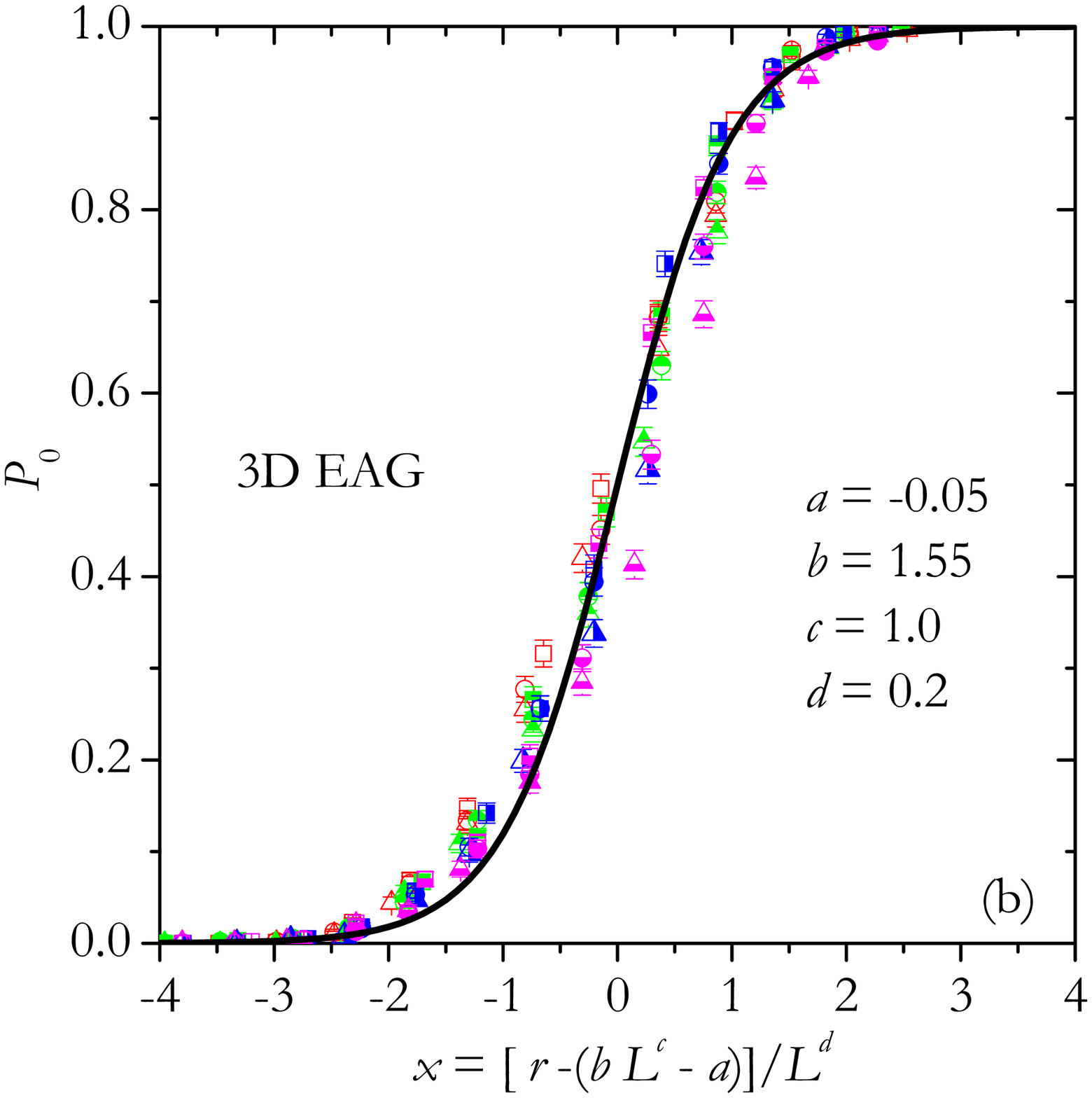}
\caption{\label{figure12}  (a) $r_0$ and $r_0^*$ vs $L$ for the 3D
EAG model.  (b) Data collapsing for curves in Fig.~\ref{figure11}
(b) between $L=5$ and $L=8$. The full line is the function defined
in Eq.~(\ref{f}) with $q=2$.}
\end{figure}

Now, we run the PT algorithm for the EAB model with $L=4$, 6, 8
($N_{\mathrm{s}}=10^3$ for each size) and $L=10$
($N_{\mathrm{s}}=10^2$). In all cases, we set $n=1$.
Figures~\ref{figure9} (a) and (b) show the mean probability $P_0$
vs $\ln(t)$ and $r$, respectively. The collapse obtained is very
good, as for the 2D EAB model. As shown in Fig.~\ref{figure10}
(a), $r_0$ is now a linear function of $L$ with $b=1.36(4)$ and
$a=1.57(33)$. If the data are rescaled using Eq.~(\ref{x}), a good
collapse is obtained, as shown in Fig.~\ref{figure10} (b) for
$c=1$ and $d=1$ (in the linear fit and data collapsing were only
considered lattice sizes $L=6$, 8 and 10). As before,
Eq.~(\ref{tm}) gives a very nice fit of all these data points, but
now with $q=6.5$.

Unfortunately, when the same analysis is carried out for the EAG
model the results are not so good. Figures~\ref{figure11} (a) and
(b) show that the collapse of the curves as function of $r$ is not
as good as for the bimodal case (even in the high $P_0$ region).
Here, the parameters used were $N_{\mathrm{s}}=10^3$ samples for
each size $L=4$, 5, 6, 7 and 8. Figure~\ref{figure12} (a) shows
the dependence of $r_0$ with $L$. Using the same fitting function,
we obtain $b=1.67(7)$ and $a=0.66(43)$ (only lattice sizes with
$L>4$ were considered).  The collapse of the data obtained with
these values is not very good, we have preferred to fit the data
for only one value of $m$ ($m=20$).  Figure~\ref{figure12} (a)
shows the dependence of $r_0^*$ ($r_0^*$ is the value of $r$ at
$P_0=0.5$ and $m=20$) on $L$, where the fit for $L>4$ gives
$b=1.55(3)$ and $a=-0.05(20)$. Data collapsing in
Fig.~\ref{figure12} (b) has been obtained with $c=1$ and $d=0.2$.
The function Eq.~(\ref{tm}) with $q=2$ gives a good fit for
$m=20$, and a reasonable good fit for all the other points. If
small lattice sizes are discarded ($L=4$ and 5), we obtain a good
collapse for $b=1.50(2)$, $a=-0.43(11)$, $c=1$, $d=1$ and $q=10$,
(these parameters are similar to the corresponding ones in the 3D
bimodal case).

\begin{figure}[t]
\includegraphics[width=8cm,clip=true]{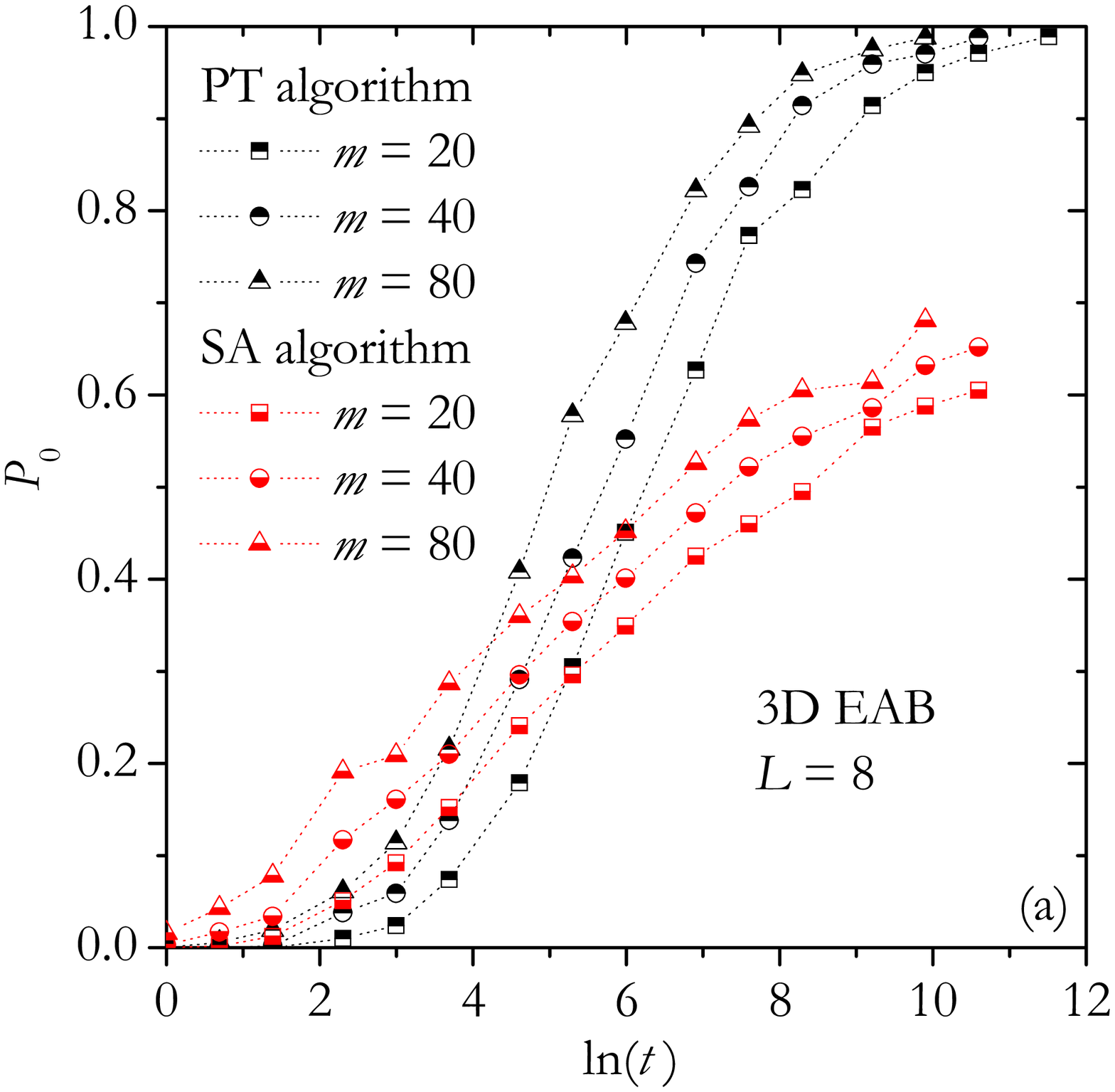}
\includegraphics[width=8cm,clip=true]{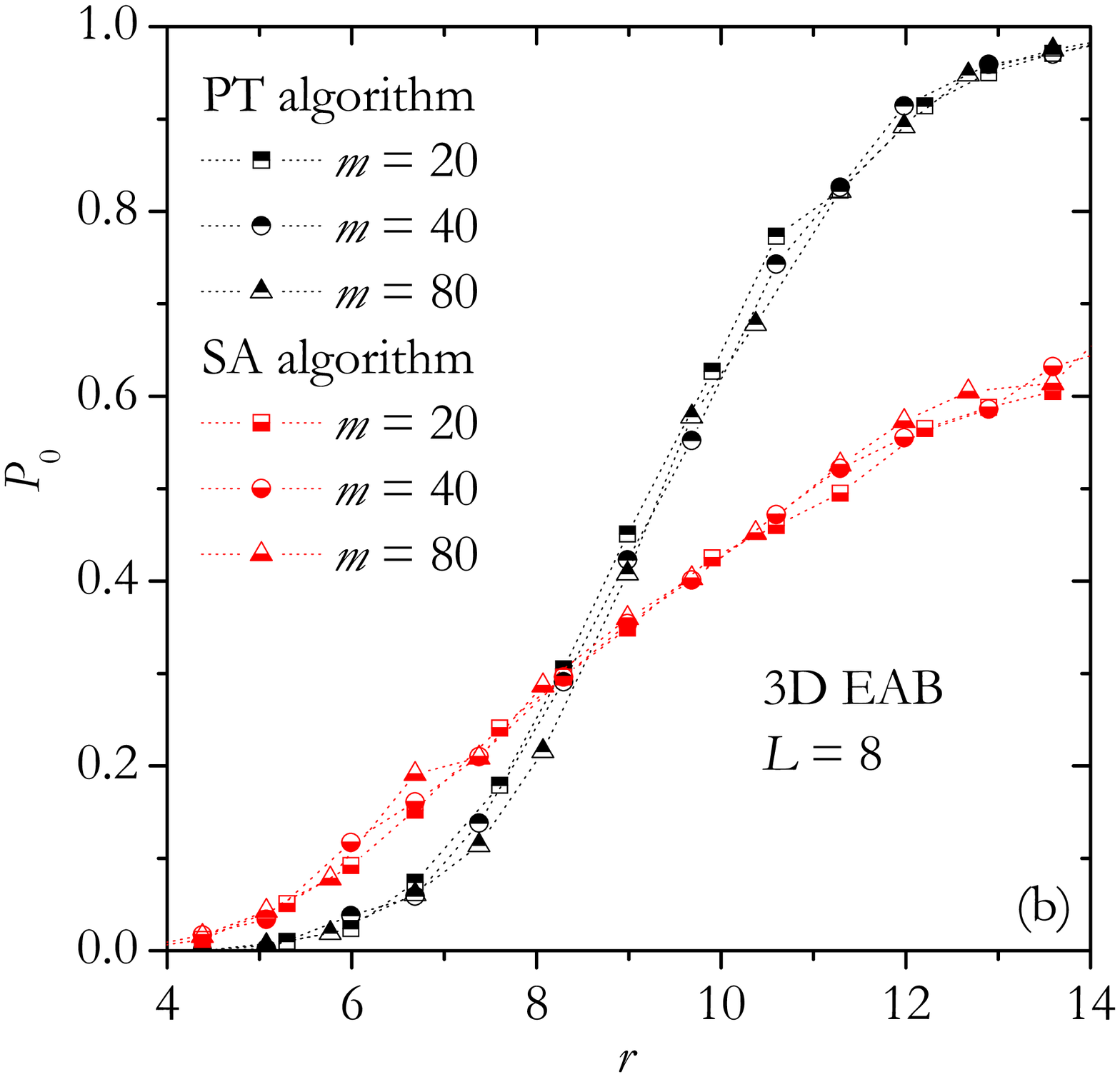}
\caption{\label{figure13}   Comparison between the PT and SA
algorithms for the 3D EAB model with $L=8$. Figures show the
probability $P_0$ as function of (a) $\ln(t)$ and (b) $r$.}
\end{figure}

Finally, to show the importance of the replica exchange procedure,
we compare the performances of the PT and the SA algorithms.  In
Figs.~\ref{figure13} (a) and (b), we show results for the 3D EAB
model with $L=8$. To carry out an appropriate comparison, we
implement the SA as the first stage of our PT algorithm A, but now
we define $t_{\mathrm{SA}}$ as the number of MCSs used in each
temperature. By choosing $t_{\mathrm{SA}}=4t$, we ensure that both
algorithms have the same running time. It is evident that the PT
algorithm is superior and only for short $t$ the SA method is more
efficient. Moreover, for this same system we have compared the
performance of both algorithms for the same samples directly, and
we have calculated the corresponding Pearson coefficient $\rho$
\cite{Feller}, to understand the correlation between them. For
each one of a set of $N_s=100$ samples and different values of
$t$, we have performed $n=100$ runs of each algorithm to obtain
$P_{0,j}$. We obtain that the Pearson coefficient of $P_{0j}$ is
very close to $1$ up to $t \approx 1000$, and then it starts to
slowly decrease. This means that for small values of $t$ the
correlation between the algorithms is almost perfect, whereas when
$t$ is further increased there are some samples which are much
"harder" for the SA.

\subsection{Ground state energy \label{sec3C}}

\begin{table}
\caption{\label{table1} Parameters of Eqs.~(\ref{tsec}) and
(\ref{tm}) for the different studied models. }
\begin{tabular}{ccccccc}
\hline
Model  &   $a$  &   $b$  &  $c$  &  $d$  &  $q$  &      $\alpha$       \\
\hline
2D EAB & $5.86$ & $3.35$ & $0.5$ & $0.2$ & $2.0$ & $1.1\times10^{-7}$  \\
2D EAG & $6.34$ & $4.61$ & $0.5$ & $0.2$ & $2.0$ & $2.1\times10^{-7}$  \\
3D EAB & $1.57$ & $1.36$ & $1.0$ & $1.0$ & $6.5$ & $1.4\times10^{-7}$  \\
3D EAG & $-0.05$& $1.55$ & $1.0$ & $0.2$ & $2.0$ & $2.7\times10^{-7}$  \\
\hline
\end{tabular}
\end{table}

We performed the calculation of the GS energy per spin for large
2D and 3D lattices.  In all cases, we chose $m=20$ and as before,
$n=1$, $T_1=1.6$ and $T_{m}=0.1$. To predict the number of PTSs
needed to obtain a given probability $P_0$, we used Eq.~(\ref{tm})
with the parameters given in Table \ref{table1}. All calculations
were carried out using a computer cluster of 40 PCs each with a
3.0 GHz Dual Intel(R) Xeon(TM) processor \cite{Nota3}. The running
time $t_{\mathrm{sec}}$ can be estimated with Eq.~(\ref{tsec}) and
the value of $\alpha$ given in Table \ref{table1}.

\begin{figure}[t]
\includegraphics[width=8cm,clip=true]{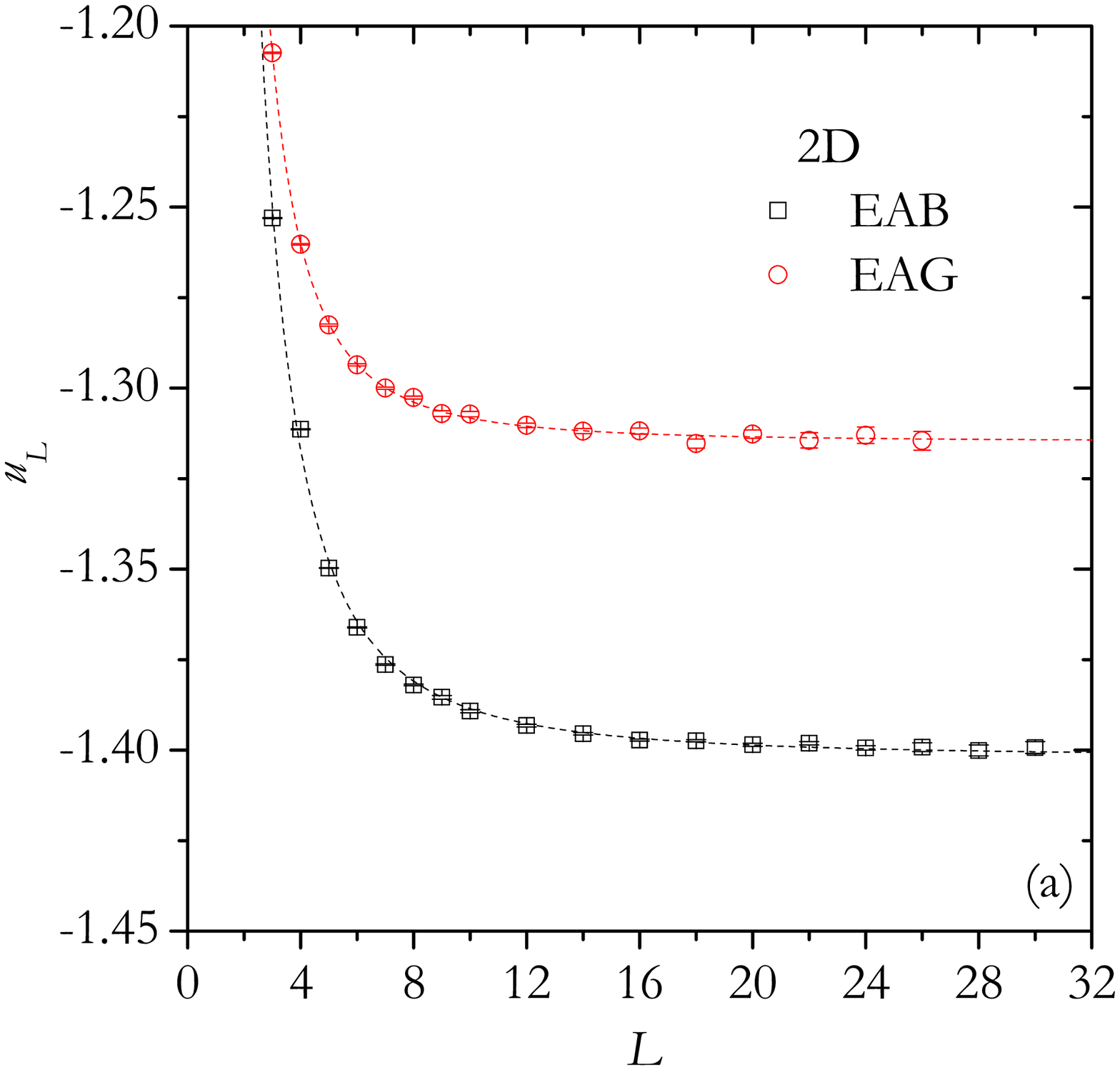}
\includegraphics[width=8cm,clip=true]{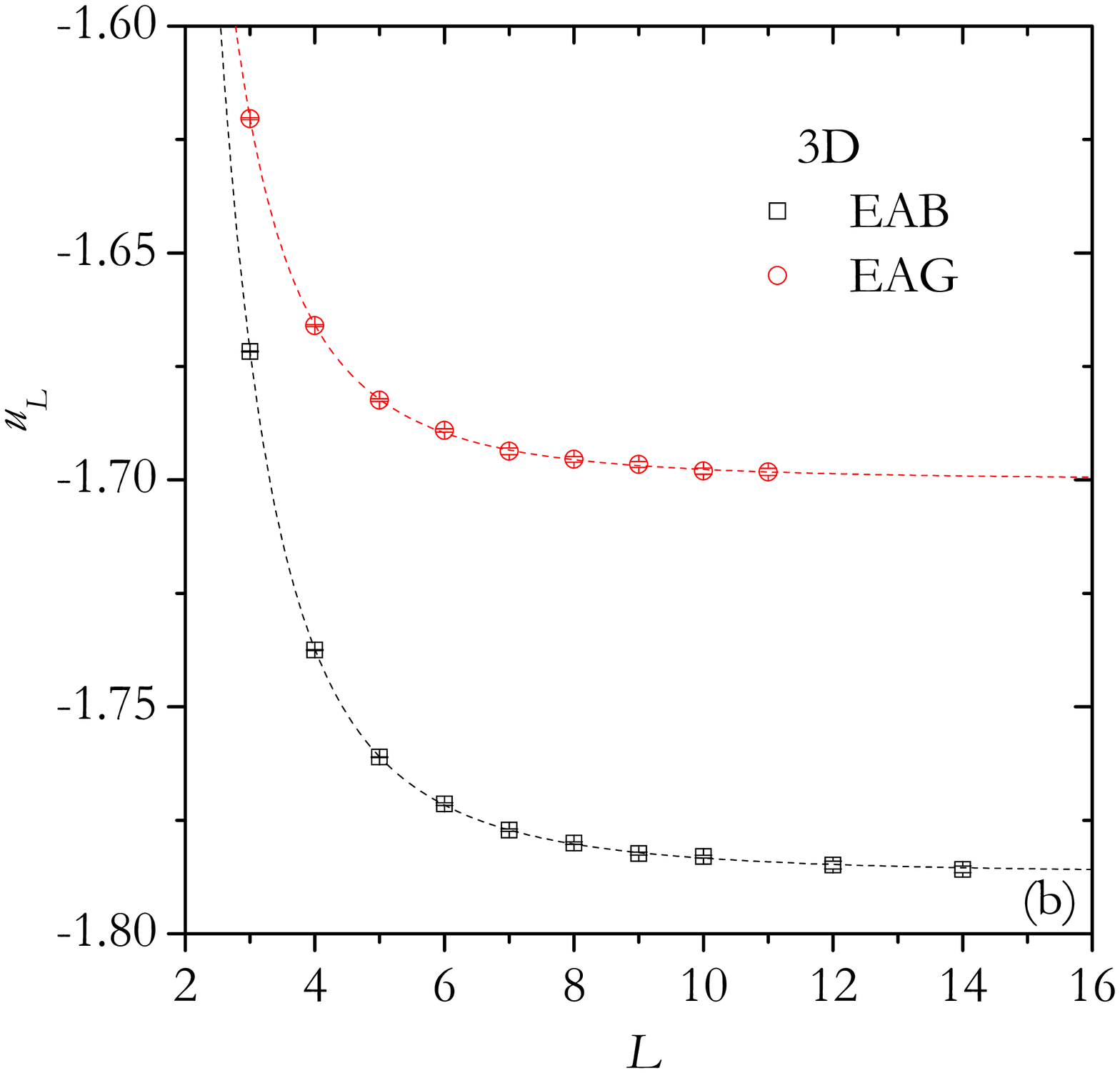}
\caption{\label{figure14}  GS energy per spin for the EAB and EAG
models in (a) 2D and (b) 3D. The dashed lines correspond to fits
with Eq.~(\ref{h1}).}
\end{figure}

Figures~\ref{figure14} (a) and (b) show the lattice size
dependence of the average of the GS energy per spin, $u_L$, for
each studied model.  The values of $u_L$, along with the
parameters used in the simulation, namely the number of samples
and the number of PTSs [calculated with Eq.~(\ref{tm})], are
summarized in Tables in appendix B.  With these quantities, the GS
is found with a mean probability $P_0$ (which is also shown in
Tables of appendix B).  For 2D lattices, we have checked that
Eq.~(\ref{tm}) remains valid (approximately within $1\%$) up to
$L=30$ for the EAB and $L=26$ for the EAG models, by running the
branch-and-cut algorithm \cite{DeSimone1995,Cologne} on the same
set of samples show in Tables \ref{table2} and \ref{table3}.
Although we cannot do the same for the 3D systems, note that for
each lattice size the values we have obtained for the GS energy in
3D, Tables \ref{table4} and \ref{table5}, agree very well with the
ones reported previously in the literature for the EAB
\cite{Pal1996a,Hartmann1997,Li2002} and EAG
\cite{Pal1996b,Palassini1999} models.

To obtain the thermodynamic limit and to understand the finite
size behavior of the GS energy, we have fitted our data in 2D and
3D with three different functional forms:
\begin{equation}
h_1=u_\infty +g L^{-e}, \label{h1}
\end{equation}
\begin{equation}
h_2=u_\infty +g L^{-D} \label{h2}
\end{equation}
and
\begin{equation}
h_3=u_\infty [1 - L^{-e}], \label{h3}
\end{equation}
where $u_\infty$ is the GS energy in the thermodynamic limit, $D$
is the lattice dimension and $g$ and $e$ are two parameters.  The
functional form of Eq.~(\ref{h2}) which only has two free
parameters, has a correction term depending of the number of
spins.  This is reasonable for 3D models where, as
$T_{\mathrm{c}}>0$, there is no scaling theory for the energy of
the GS.  For 2D models, however, the leading finite size term is
predicted to be $L^{-(D-\theta)}$ \cite{Campbell2004}, where
$\theta$ is the stiffness exponent which characterizes the scaling
of the domain walls energy. Therefore, Eq.~(\ref{h2}) is
equivalent to assuming $\theta=0$, and in 2D it should give a good
fit for the EAB model but not for the EAG model (see below).  The
other functional form with two free parameters, Eq.~(\ref{h3}),
has been chosen because it gives a `reasonable' limit of
Eq.~(\ref{h1}) for $L=1$: $u_L=0$ at $L=1$ \cite{Campbell2004}.

The parameters giving the best fits for the three functional
forms, and the corresponding goodness-of-fit parameter $Q$
\cite{numerical} are given in Table \ref{table1b}. A value $Q>0.1$
is usually considered as indication of good quality of the fit. As
is usually the case, scaling functions do not include all possible
finite size corrections, and therefore better fits are obtained
when data for very small sizes are left out. On the other hand,
leaving out too many points can result in large error bars for the
best fit parameters. The results we show were obtained by fitting
the data over the largest range that gives a goodness-of-fit of $Q
\gtrsim 0.1$. This is important because large error bars in the
fitting parameters could mask the differences between the
functional forms proposed.

\begin{table}
\caption{\label{table1b} Best fit parameters for the three scaling
functions tested [Eqs.~(\ref{h1}), (\ref{h2}) and (\ref{h3})]. $Q$
is a measure of the goodness of fit.}
\begin{tabular}{ccccccc}
\hline
Model  & Function &  $u_\infty$  &   $g$     &   $e$     &  $Q$   & Range of $L$    \\
\hline
2D EAB &  $h_1$   & $-1.4009(3)$ & $1.6(1)$  & $2.13(4)$ & $0.15$ &  $5-30$ \\
       &  $h_2$   & $-1.4013(2)$ & $1.23(2)$ &           & $0.20$ &  $7-30$ \\
       &  $h_3$   & $-1.4011(2)$ &           & $2.07(1)$ & $0.22$ &  $7-30$ \\
2D EAG &  $h_1$   & $-1.3149(5)$ & $1.3(2)$  & $2.28(9)$ & $0.10$  &  $4-26$ \\
       &  $h_2$   & $-1.3156(4)$ & $0.78(3)$ &           & $0.32$ &  $6-26$ \\
       &  $h_3$   & $-1.3149(3)$ &           & $2.30(1)$ & $0.17$ &  $4-26$ \\
3D EAB &  $h_1$   & $-1.7867(2)$ & $2.89(6)$ & $2.93(2)$ & $0.29$ &  $3-14$ \\
       &  $h_2$   & $-1.7866(2)$ & $3.21(4)$ &           & $0.14$ &  $5-14$ \\
       &  $h_3$   & $-1.7875(2)$ &           & $2.64(1)$ & $0.62$ &  $6-14$ \\
3D EAG &  $h_1$   & $-1.7000(3)$ & $2.01(8)$ & $2.94(4)$ & $0.80$ &  $3-11$ \\
       &  $h_2$   & $-1.6997(2)$ & $2.14(1)$ &           & $0.67$ &  $3-11$ \\
       &  $h_3$   & $-1.7004(2)$ &           & $2.82(1)$ & $0.84$ &  $4-11$ \\
\hline
\end{tabular}
\end{table}

For the 2D EAB model we obtain good fits for $h_1$ in the range
$5-30$, whereas a smaller number points is needed to obtain a
comparable $Q$ for $h_1$ and $h_2$. The thermodynamic limit
obtained with the three functions agree within error bars, but
they are slightly different to the most accurate value reported,
$u_\infty=-1.40193(2)$ \cite{Palmer1999}. If the fit is performed
over a larger interval, namely for $3-30$, the value obtained
$u_\infty=-1.4019(8)$ agrees with the one given by Palmer and
Adler, but the goodness of fit is very bad.

For the exponent of the correction term, the situation is less
clear. Fixing the exponent to $2$ (i.e. choosing $h_2$) gives a
very good fit to the data, whereas using this exponent as a
fitting parameter gives a fit almost as good, but for larger
values: $e=2.13(4)$ for $h_1$ and $2.07(1)$ for $h_3$. Exponents
closer to $2$ can be obtained by fitting over a larger range
[$e=2.02(5)$ for $h_1$ in the range $3-30$, and $e=2.01(1)$ for
$h_3$ in the range $4-30$] but with a vanishing goodness of fit.

If only one value has to be chosen, it should be $e=2.13(4)$
because it has been obtained by fitting over the largest range.
This value, however, is a bit larger than expected. As mentioned
above, scaling theory predicts that this exponent should be equal
to $D-\theta$. By measuring the domain wall energy, Hartmann and
Young have obtained that $\theta=0$ \cite{Hartmann2001}. The
larger value we obtain might be due to an additional correction
term whose exponent is very similar to the one predicted by
scaling theory. In this case the sum of such terms would look like
a single correction term with an `effective' $\theta$, as it has
been suggested in Ref.~\cite{Campbell2004}.

For the 2D EAG model, our results for $h_1$ and $h_3$ are in good
agreement with the scaling theory prediction, using $\theta =
-0.287(4)$ \cite{Hartmann2002}. For these functional forms the
thermodynamic limits of the GS energy also agree very well with
the most accurate report, $u_\infty=-1.31479(2)$
\cite{Campbell2004}.  On the other hand, the result obtained with
$h_2$ is consistent with this: a good fit is achieved only for
smaller range of $L$ given $u_\infty=-1.3156(4)$, which does not
agree, within error bars, with the value reported by Campbell {\em
et al.} \cite{Campbell2004}.

For 3D models the number of sample sizes available is so small
that it is usually not easy to decide which functional form gives
a better fit of the data. In this case one is forced to choose,
among fits of similar quality, those that span the largest range.
It must be stressed, however, that this choice hinges usually on
only a couple of points, and therefore it is not improbable that
simulations for larger sizes could tip the scale in favor of the
other functional forms. For example, for the 3D EAB model one
should choose the functional form $h_1$. The energy and exponent
obtained (see Table \ref{table1b}) agree very well with the values
found in Ref.~\cite{Pal1996a}, $u_\infty=-1.7863(4)$ and
$e=2.965(11)$, and are very close to reported in
Ref.~\cite{Hartmann1997}, $u_\infty=-1.7876(3)$ and $e=2.84(5)$.
On the other hand, it has been suggested \cite{Bouchaud2003} that,
even though $T_{\mathrm{c}}>0$, for 3D the finite size dependence
of the GS energy could follow the same scaling law as for 2D.
However, the exponent given by $h_1$ is not consistent with this
prediction and with the fact that the value of the stiffness
reported is $\theta=0.19(2)$ \cite{Hartmann1999}, unless there are
additional scaling corrections that give rise to an effective
$\theta$.

For the 3D EAG model we find that the fits given by $h_1$ and
$h_2$ are equivalent, and thus the possibility that the exponent
be exactly $3$ cannot be ruled out.  Nevertheless, it is
interesting to see that the exponent obtained with $h_1$ is very
similar to the one obtained with this same function for the
bimodal case. For the 3D EAG model, however, the stiffness
exponent has been found to be $\theta=0.24(1)$
\cite{Boettcher2001}, making it less likely that the scaling
hypothesis can be applied to this model. In addition, the energy
values obtained agree very well with the value reported by P\'al,
$u_\infty=-1.7003(8) $ \cite{Pal1996b}.

We have tried to fit our data with functional forms that depend
exponentially on the sample size. The qualities of the fits are in
general very poor both in 2D and 3D, so that they can be ruled out
as candidates for scaling functions for the data, as has been
previously reported \cite{Pal1996a,Hartmann1997}.

Finally, for the bimodal cases we have also performed simulations
over systems where the bonds are not chosen so as to enforce
Eq.~(\ref{constraint}), i.e. their sign is instead chosen
independently and with equal probabilities.  For each size we have
performed simulations over the same number of samples as in the
`constrained' case.  For $L \geq 5$ in 2D and $L \geq 3$ in 3D,
the GS energies we obtain are not statistically different to those
shown in Tables \ref{table2} and \ref{table4}, and the parameters
giving the best fits are in good agreement with the values
reported in Table \ref{table1b}. However, the goodness of fit is
certainly not as good, which is probably due to the `noise'
introduced by the fluctuations in the number of bonds of each sign
\cite{Pazmandi1997}.

\section{Probabilistics of failures \label{sec4}}

In the previous section we have found a simple expression,
Eq.~(\ref{tm}), that is useful to estimate the approximate number
of PTSs that the PT algorithm needs to find a GS with a given mean
probability $P_0$.  The accuracy of this equation is enough to
calculate the average of the GS energy per spin as the previous
section, for all practical purposes. However, Eq.~(\ref{tm})
cannot be used to predict the probability of reaching the GS for a
single sample. As we have previously observed for 3D systems,
there exist hard (easy) samples for which, for a given $t$, the GS
is found with a probability smaller (bigger) that $P_0$. This
could in turn distort the sample average, if the hard samples were
numerous enough, or if the variable to be calculated is strongly
correlated with the hardness of the sample.

The best way to address these issues would be to build a histogram
of the minimal $t$ needed to reach the GS of a large number of
samples, for a fixed probability. Unfortunately, this would demand
a huge computational effort as for each sample a huge number of
runs would be needed to determine the histogram. For this reason,
in the following we use an indirect method applied recently by
Weigel \cite{Weigel2007} to study the performance of a genetic
embedding matching algorithm. We applied it to the analysis of the
3D EAB model, for $L=6$.

\begin{figure}[t]
\includegraphics[width=8cm,clip=true]{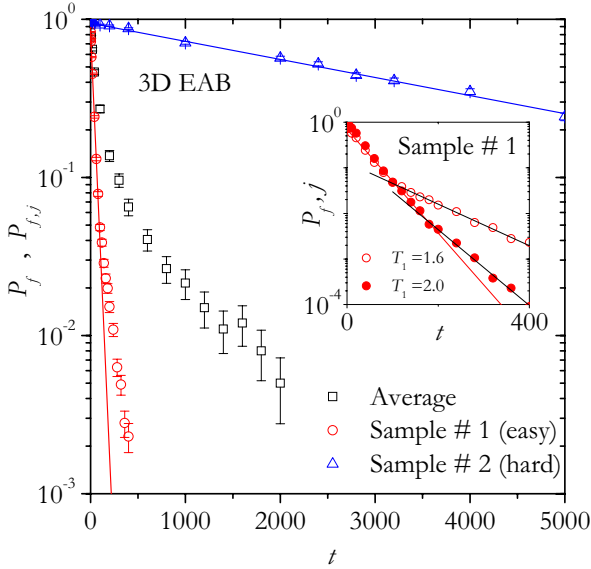}
\caption{\label{figure15}  Failure probability as a function of
$t$, for two samples (easy and hard) of the 3D EAB model with
$L=6$, and the average over $10^3$ samples. The lines are linear
best fits.}
\end{figure}

For each one of the $10^3$ samples we have calculated the failure
probability $P_{f,j}=1-P_{0,j}$ for different values of $t$ (a
number of up to $n=10^4$ independent runs for sample were carried
out).  The parameters used were the same as in the previous
section: $m=20$, $T_1=1.6$ and $T_{m}=0.1$). Figure~\ref{figure15}
shows the failure probabilities for two different samples, the
same that in the previous section were called easy and hard. It
can be seen that for small number of PTSs the data for each sample
are well approximated by an exponential function
\begin{equation}
P_{f,j} (t)=K_j^t . \label{pf}
\end{equation}
The constant $K_j$ quantifies the hardness of each sample. Notice
that this relation implies that the algorithm is quite efficient,
in the sense that a run of a given number of steps has the same
probability of finding a GS as two independent runs with half the
number of PTSs. As is to be expected, Fig.~\ref{figure15} shows
that this does not hold for the sample average of the failure
probability, $P_f$. On the other hand, for single samples we have
found that, after a certain amount sample-dependent time, the
probability of failure decays more slowly. This is an indication
that in some sense, the algorithm loses efficiency {\em for a
given set of parameters}. As the inset of Fig.~\ref{figure15}
shows, increasing only the highest temperature $T_1$ allows the
algorithm to operate efficiently for larger $t$. Notice that, as
explained before, another way to stay in the efficient regime is
simply to split the desired number of PTSs into several
independent runs.

\begin{figure}[t]
\includegraphics[width=8cm,clip=true]{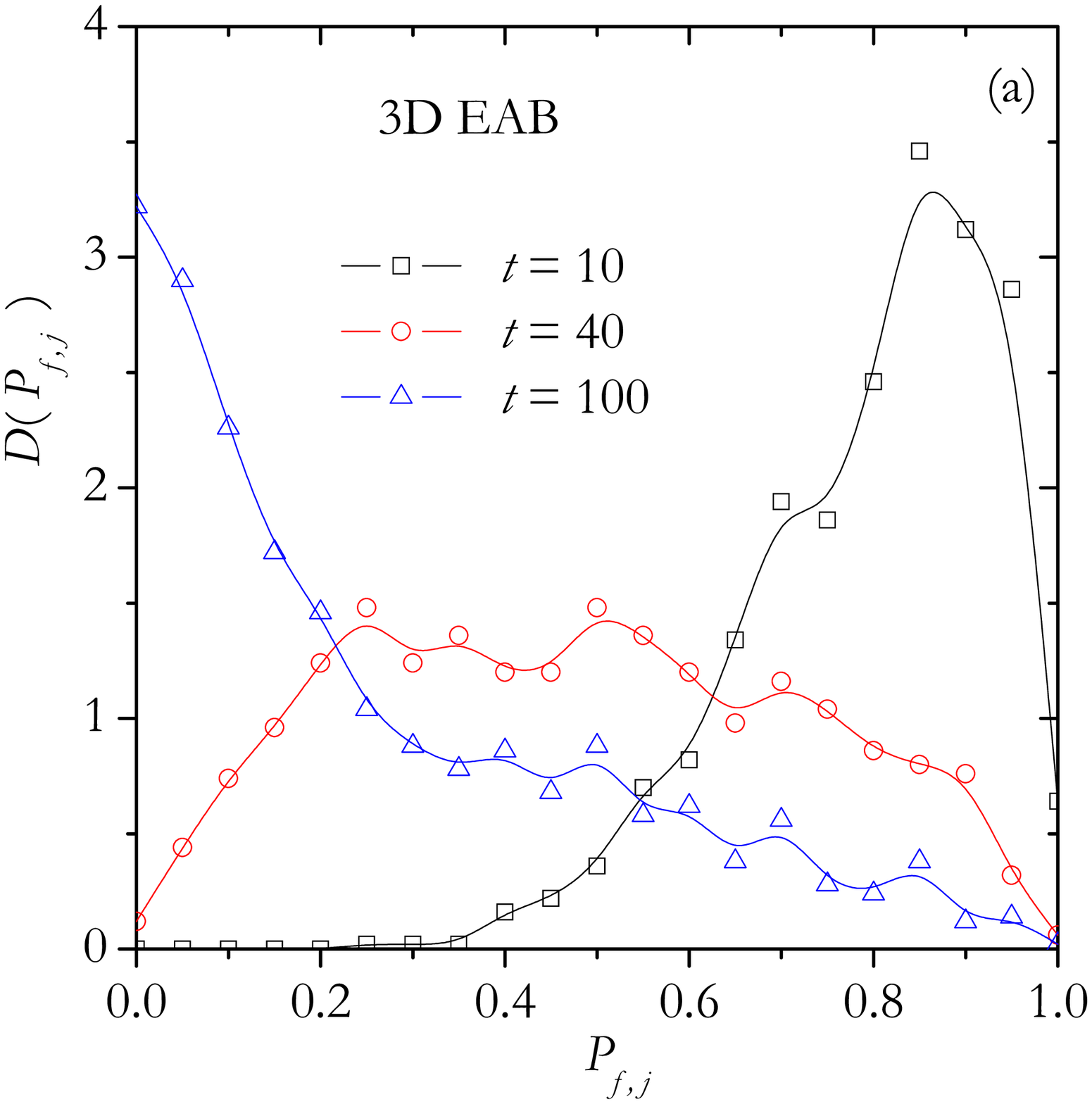}
\includegraphics[width=8cm,clip=true]{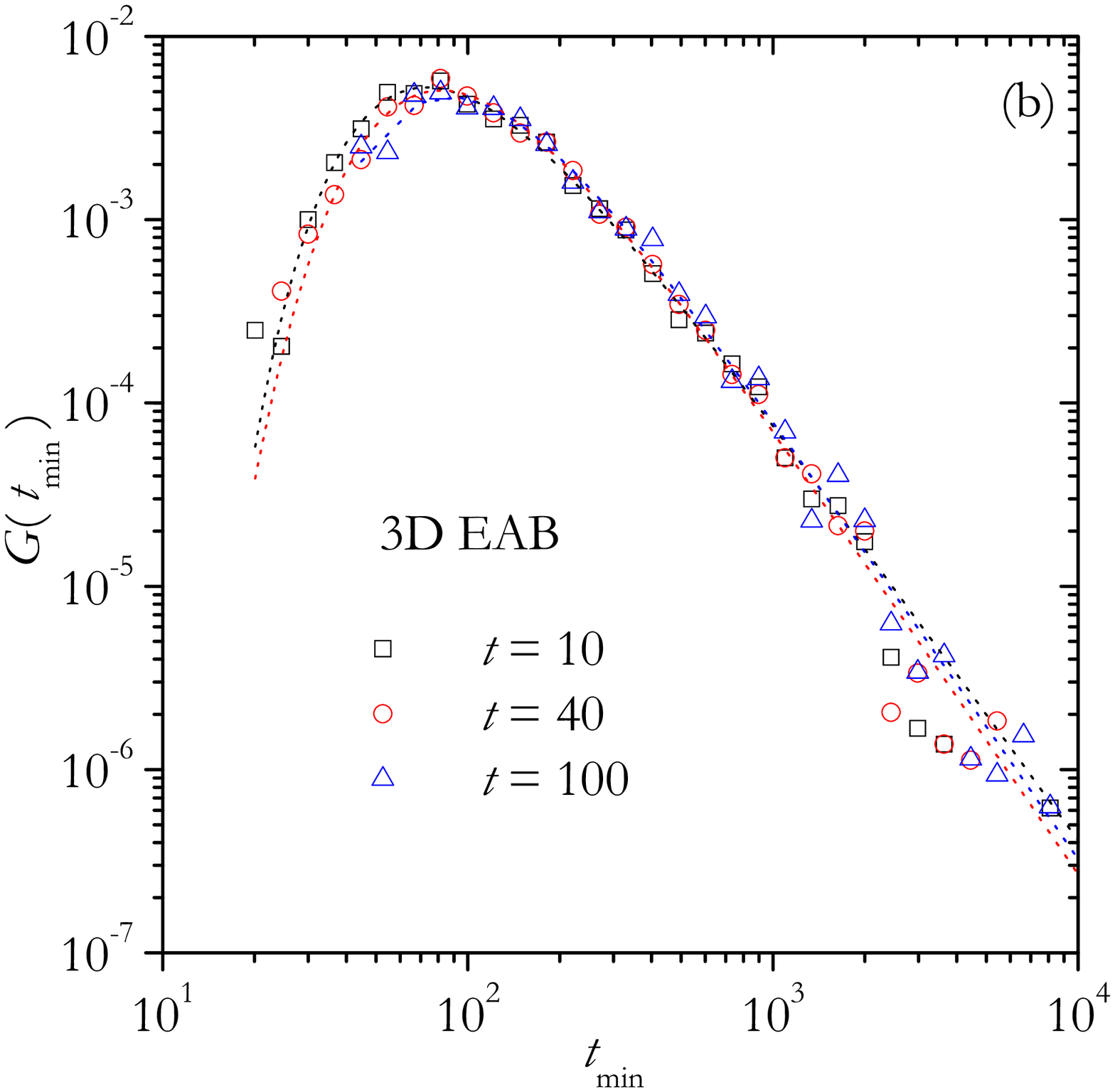}
\caption{\label{figure16}  (a) Histograms for the failure
probability at three different $t$. The full lines in are only
guides to the eye.  (b) Distribution of minimum required number of
PTSs needed to achieve a given failure probability $P_f^*=0.05$.
The dotted lines are best fits of a Fr\'echet distribution. }
\end{figure}

Figure~\ref{figure16}(a) shows $D(P_{f,j})$, the distribution of
$P_{f,j}$ for three different values of $t$. These times are small
so as to be sure to be in the regime where the Eq.~(\ref{pf}) is
valid.  Notice that the distributions are rather different: they
have peaks for high and small probabilities for $t=10$ and
$t=100$, but for an intermediate time, $t=40$, the distribution is
rather wide.

Using any one of these distributions and Eq.~(\ref{pf}), it is
possible to obtain $G(t_{\mathrm{min}})$, the distribution of the
minimum required number of PTSs, $t_{\mathrm{min}}$, needed to
achieve a given failure probability $P_f^*$.  In fact, from the
estimate $P_{f,j}(t)$ at fixed $t$ for a given sample,
Eq.~(\ref{pf}) implies
\begin{equation}
t_{\mathrm{min}}=t \frac{\ln P_f^*}{\ln P_{f,j}(t)}. \label{tmin}
\end{equation}
Figure~\ref{figure16}(b) shows for $P_f^*=0.05$, the function
$G(t_{\mathrm{min}})$ for times $t=10$, $40$ and $100$.  Notice
that the distributions obtained are very similar, which confirms
that the algorithm was indeed in the efficient regime.

\begin{figure}[t]
\includegraphics[width=8cm,clip=true]{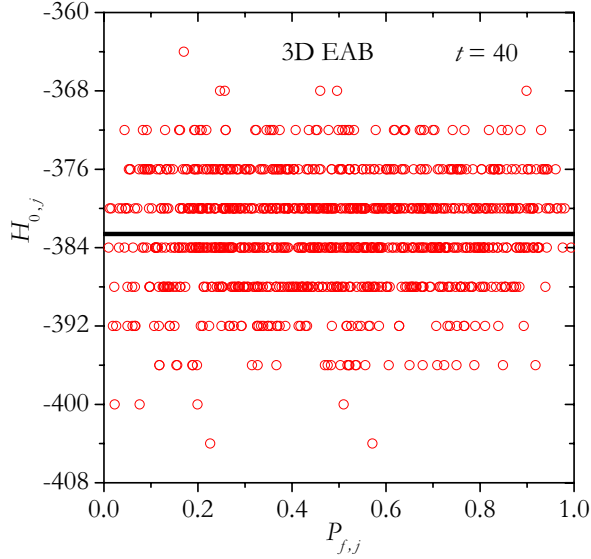}
\caption{\label{figure17}  Exact GS energy as a function of the
failure probability that the GS is found in $t=40$ steps, for
$10^3$ samples of the 3D EAB model with $L=6$. The line represent
the average energy of all the samples.}
\end{figure}

As in Ref.~\cite{Weigel2007}, we found a very wide distribution
$G(t_{\mathrm{min}})$ with a fat tail. Extremal value theory
predicts that the distribution $G(t_{\mathrm{min}})$ should be
given by a function of the form \cite{Beirlant}
\begin{equation}
G_{\xi, \mu, \sigma}(x) = \frac{1}{\sigma} \left( 1+\xi
\frac{x-\mu}{\sigma} \right)^{-1-1/\xi} \exp \left[ -\left( 1+\xi
\frac{x-\mu}{\sigma} \right)^{-1/\xi} \right], \label{Frechet}
\end{equation}
where $\sigma$, $\mu$ and $\xi$ are three parameters.
Figure~\ref{figure16}(b) shows that this function fits our data
very well (i.e. with a high quality-of-fit $Q$).  For times
$t=10$, $40$ and $100$ we obtain, respectively: $\sigma=89(3)$,
$\mu=115(2)$ and $\xi=0.78(5)$ with $Q=0.28$; $\sigma=87(4)$,
$\mu=121(2)$ and $\xi=0.69(6)$ with $Q=0.53$; $\sigma=96(5)$,
$\mu=129(3)$ and $\xi=0.71(9)$ with $Q=0.27$.  The values of the
$\xi$ indicate that this is a distribution of the Fr\'echet type
($\xi > 0$) with a divergent variance ($\xi > 1/2$)
\cite{Beirlant}.  This means that the number of hard samples is
not negligible. Nevertheless, we find no sign of correlation
between the exact GS energy of a given sample, $H_{0,j}$, and the
failure probability that their GS is found for the PT algorithm.
For example, Fig.~\ref{figure17} shows these quantities for $10^3$
samples of the 3D EAB model with $L=6$.  The average GS energy of
this set is $H_0 = -382.6(2)$, but it is $H_0 = -382.7(3)$ or $H_0
= -382.6(3)$ if, respectively, samples with $P_{f,j} < 1/2$ or
$P_{f,j} > 1/2$ are chosen to calculate the mean value.  As we can
seen, the three average GS energy values agree within statistical
bounds. In addition, the Pearson correlation coefficient $\rho$
\cite{Feller} between $H_{0,j}$ and $P_{f,j}$ is $\rho =
0.037(32)$, and between $H_{0,j}$ and $K_j$ (the hardness of each
sample) is $\rho = 0.036(34)$.  Although a coefficient $\rho
\approx 0$ does not assure independence, it is evident from
Fig.~\ref{figure17} that this is the most probable scenario.  We
conclude that although the $G(t_{\mathrm{min}})$ is a Fr\'echet
distribution with a divergent variance, the calculation of the
average GS energy will not be affected for this reason, because
the hardness of each sample (under the PT algorithm) is
independent of their own GS energy.

\section{Conclusions \label{sec5}}

In this work we have used a PT algorithm to find the GS energy of
the EA model with both bimodal and Gaussian bond distributions. In
general, this heuristic can be easily implemented to solve a very
general class of problems: systems with any boundary conditions,
with arbitrary forms of interactions, with or without external
field, with any dimensionality, etc. This is the most important
feature of the PT algorithm. We have checked that, for large
lattice sizes, the A variant of our algorithm is always a lower
bound of the performance of variants B and C in the high $P_0$
regime ($P_0>0.9$). For practical purposes, Eq.~(\ref{tm}) with
the parameters given in Table 1 can be used to calculate the
number of PTSs for the three variants (with $t_A=2t$, $t_B=2.3t$
and $t_C=1.5t$). For a given $t$, the error in the probability
$P_0$ predicted by Eq.~(\ref{tm}) is not larger than $1\%$.

The performance of PT is comparable to the performance of more
powerful heuristics, developed exclusively to find the GS of Ising
spin glass systems. In 2D, this algorithm allows us to study
systems with lattice sizes up to approximately $L = 30$ and $L =
26$ for, respectively, EAB and EAG models with fully periodic
boundary conditions. Although larger sizes can be analyzed by
using matching algorithms \cite{Hartmannlibro1}, this can only be
done for planar lattices (i.e. lattices with at least one free
boundary condition). But, for such lattices, it has been found
that very large system sizes must be used to have a reliable
estimate of the thermodynamic limit of the GS energy (and other
quantities), which somewhat undermines the advantages of having a
faster algorithm. For lattices with fully periodic conditions, on
the other hand, it has been shown that the energy converges to the
thermodynamic limit for relatively small system sizes. For these
systems, the branch-and-cut algorithm \cite{DeSimone1995} is the
fastest exact algorithm to calculate GSs of the EA model.
Unfortunately its implementation is rather difficult. As far as we
know, the most efficient implementation of this heuristic is
available on the server at the University of Cologne
\cite{Cologne}. On the other hand, recently it has been shown that
a {\it patchwork dynamics} can be used to calculate correct GSs of
samples with large lattices sizes \cite{Thomas2008}.  This
technique seems to be very promising to study even 3D spin
glasses. Lastly, we note that a new MC algorithm (a high-efficient
PT algorithm) could be used as an efficient heuristic to calculate
GSs, because this technique allows to equilibrate 2D systems of
size $L=10^2$ down to temperature $T=0.1$ \cite{Houdayer2001b}.

Contrary to the 2D case, finding the GS of a spin glass in a 3D
lattice is a very difficult task, which has even been shown to be
NP- complete \cite{Barahona1982}. Although an exact branch-and-cut
algorithm has been developed for the EAG model, it can only find
GSs of samples up to $L = 12$ with free boundary conditions
\cite{Palassini2003,Liers2000}. Thus, 3D systems with fully
periodic boundary conditions constitute the most important
application of heuristic algorithms. Among these, genetic search
methods are usually considered as the most powerful techniques to
find the GS of spin glass systems. Nevertheless, in this work we
have shown that, for the same task, a simple PT algorithm performs
as well as the genetic methods reported in the literature (i.e.
similar systems sizes can be analyzed with the same computational
effort). For example, for the 3D EAB model it has been reported
that a genetic algorithm needs on average 392 minutes on a
computer with a 134MHz R4600 processor \cite{Pal1996a}, or 540
minutes on a computer with a 80MHz PPC601 processor
\cite{Hartmann1997}, to perform a run in samples of $L = 14$. In
Ref.~\cite{Hartmann1997}, $10^2$ samples of this size were
calculated and $40$ independent runs for each sample were carried
out. On average, in only $13.8$ of these runs the lowest energy
was obtained \cite{Hartmann1997}. Thus, we deduce that the GS was
found with $P_0 \approx 13.8/40=0.345$. For this same probability,
our algorithm requires approximately $60$ minutes on a computer
with a $3.0$ GHz Dual Intel(R) Xeon(TM) processor. Although a
direct comparison between these results is inappropriate (because
these works were carried out more than ten years ago), probably
the performances of both heuristics are comparable.

On the other hand, we consider the Ref.~\cite{Pal1996b} where
samples of size up to $L = 10$ for the 3D EAG model were
calculated with $P_0 \approx 0.9$, by using a genetic algorithm
with local optimization.  With the PT algorithm, we have obtained
similar results for $L=10$, and for $L=11$ we have found the GS
with $P_0 \approx 0.8$.  On the other hand, for the same model
recently a genetic renormalization algorithm has been introduced,
which is able to solve lattices up to $L=12$
\cite{Hartmannlibro2,Houdayer2001}. Unfortunately, we have not
been able to compare our results with those obtained by using this
fast algorithm, because in Ref.~\cite{Houdayer2001} the energies
for each lattice size have not been reported.

Also, we have used the PT algorithm to study the finite size
behavior of the GS energy.  While our results for 2D EAG model are
in good agreement with the scaling theory prediction
\cite{Campbell2004}, we have not been able to prove this for the
2D EAB model (presumably due to an additional correction term with
exponent $2$).  In 3D we obtain similar exponents for EAB and EAG
models, which are not equal to $D - \theta$ as it has been
suggested in Ref.~\cite{Bouchaud2003}.  Nevertheless, it is
possible that this conjecture can be tested if larger lattices
sizes can be calculated.  In addition, the thermodynamic limit of
the GS energy obtained in all cases (2D and 3D models) agree very
well with values reported previously in the literature.

Finally, the efficiency of the PT technique to find the GS of
single samples has been studied for the 3D EAB model.  Using a
indirect method, we have shown that the minimum required number of
PTSs needed to achieve a given failure probability, follows a
Fr\'echet distribution with a divergent variance. In principle,
this could distort the sample average of GS properties, if there
was a correlation between the hardness of the sample and the
quantity to be measured. For the GS energy, however, we have found
no sign of such correlations.

\subsection*{Acknowledgments}

Three of the authors (AR-P,FN and EEV) thank Fondecyt (Chile)for
partial support under projects 1060317 and 7080020. One author
(EEV) is grateful to Millennium Scientific Initiative (Chile) for
partial support under contract "Basic and Applied Magnetism"
P06-022F.  FR, FN and AR-P thank Universidad Nacional de San Luis
and CONICET (Argentina) for partial support under projects 322000
and PIP6294, respectively.


\appendix

\section{Appendix A: Error bars}

The mean probability of reaching the GS is
\begin{equation}
\mathcal{P}_0 \equiv \langle P_{0,j} \rangle , \label{A1}
\end{equation}
where $\langle .... \rangle$ represent a sample average.  In order
to estimate $\mathcal{P}_0$, we use Eqs.~(\ref{P0}) and
(\ref{P0j}),
\begin{equation}
P_0=\frac{1}{N_{\mathrm{s}}}\sum_{j=1}^{N_{\mathrm{s}}}
\frac{n_j}{n}.
 \label{A2}
\end{equation}
In the last equation we have estimate the probability of reaching
the GS for the $j$-th sample, $\mathcal{P}_{0,j}$, with
$P_{0,j}=n_j / n$, where $n_j$ is the number of times that GS is
found for the $j$-th sample in $n$ independent runs.  In the
following, we estimate the error associated to $P_0$.

We begin by considering the mean number of times that GS is
reached for the $j$-th sample in $n$ independent runs,
\begin{equation}
\overline{n_j}=\sum_{n_j=0}^{n} n_j ~ {n \choose n_j}
\mathcal{P}_{0,j}^{n_j} (1-\mathcal{P}_{0,j})^{n-n_j}=n
\mathcal{P}_{0,j}.
 \label{A3}
\end{equation}
In addition, the variance is
\begin{equation}
V(n_j)= \overline{(n_j- \overline{n_j})^2}=n \mathcal{P}_{0,j}
(1-\mathcal{P}_{0,j}).\label{A4}
\end{equation}
Now, the error associated to $P_0$ can be estimated by calculating
the variance of Eq.~(\ref{A2}),
\begin{equation}
V(P_0)=E \left[ \left( P_0-E[P_0] \right) ^2 \right] ,\label{A5}
\end{equation}
where the expected value of any quantity $x$ is obtained as
$E[x]=\langle \overline{x} \rangle $.  The Eq.~(\ref{A5}) can be
rewritten as
\begin{eqnarray}
V(P_0) & = & E \left[\left(\frac{1}{N_{\mathrm{s}}
n}\sum_{j=1}^{N_{\mathrm{s}}}
n_j-\frac{1}{N_{\mathrm{s}}}\sum_{j=1}^{N_{\mathrm{s}}}
\mathcal{P}_0 \right) ^2 \right] \nonumber\\
& = & E \left[\left(\frac{1}{N_{\mathrm{s}}
n}\sum_{j=1}^{N_{\mathrm{s}}} n_j - \frac{1}{N_{\mathrm{s}} n}
\sum_{j=1}^{N_{\mathrm{s}}} n \mathcal{P}_{0,j} +
\frac{1}{N_{\mathrm{s}}} \sum_{j=1}^{N_{\mathrm{s}}}
\mathcal{P}_{0,j} - \frac{1}{N_{\mathrm{s}}}
\sum_{j=1}^{N_{\mathrm{s}}} \mathcal{P}_0 \right) ^2 \right]
\nonumber\\
& = & \frac{1}{(N_{\mathrm{s}} n)^2} \sum_{j=1}^{N_{\mathrm{s}}} E
\left[ \left( n_j - n \mathcal{P}_{0,j} \right) ^2 \right] +
\frac{1}{(N_{\mathrm{s}})^2} \sum_{j=1}^{N_{\mathrm{s}}} E \left[
\left(\mathcal{P}_{0,j} - \mathcal{P}_0 \right) ^2 \right] +
\nonumber\\
& & + \frac{1}{(N_{\mathrm{s}})^2 n} \sum_{j>k} E \left[ n_j - n
\mathcal{P}_{0,j} \right] E \left[ \mathcal{P}_{0,k} -
\mathcal{P}_0 \right].
\end{eqnarray}
As the last term disappears, we obtain
\begin{eqnarray}
V(P_0) & = & \frac{1}{(N_{\mathrm{s}} n)^2}
\sum_{j=1}^{N_{\mathrm{s}}} \langle n \mathcal{P}_{0,j} \left( 1-
\mathcal{P}_{0,j} \right) \rangle + \frac{\langle
\left(\mathcal{P}_{0,j} - \mathcal{P}_0
\right) ^2 \rangle}{N_{\mathrm{s}}} \nonumber\\
& = & \frac{\langle \mathcal{P}_{0,j} \left( 1- \mathcal{P}_{0,j}
\right) \rangle}{N_{\mathrm{s}} n} + \frac{\langle
\mathcal{P}_{0,j}^2
\rangle - \langle \mathcal{P}_{0,j} \rangle ^2}{N_{\mathrm{s}}} \nonumber\\
& = & \frac{\langle \mathcal{P}_{0,j} \rangle - \langle
\mathcal{P}_{0,j} ^2 \rangle}{N_{\mathrm{s}} n} + \frac{\langle
\mathcal{P}_{0,j}^2 \rangle - \langle \mathcal{P}_{0,j} \rangle
^2}{N_{\mathrm{s}}}.
\end{eqnarray}
Then, the error associated to $P_0$ is
\begin{equation}
\sqrt{V(P_0)}  = \sqrt{\frac{\langle \mathcal{P}_{0,j} \rangle -
\langle \mathcal{P}_{0,j} ^2 \rangle}{N_{\mathrm{s}} n} +
\frac{\langle \mathcal{P}_{0,j}^2 \rangle - \langle
\mathcal{P}_{0,j} \rangle ^2}{N_{\mathrm{s}}}}.
\end{equation}
Note that it is not necessary to consider many runs for sample:
the error becomes small if many samples are considered and only
one run is carried out in each one of them. Considering that $n=1$
and $P_0 \approx \langle \mathcal{P}_{0,j} \rangle $, we approach
the error of $P_0$ by
\begin{equation}
\sqrt{V(P_0)} \approx \sqrt{ \frac{P_0 ( 1-
P_0)}{N_{\mathrm{s}}}}.
\end{equation}

\section{Appendix B: Tables}

Parameters used in the simulation and GS energy per spin for each
lattice size.

\begin{table}
\caption{\label{table2} Simulation parameters and GS energy per
spin for the 2D EAB model. }
\begin{tabular}{ccccc}
\hline
$L$&     $u_L$    &     $N_{\mathrm{s}}$     &      $t$      &$P_0$  \\
\hline
2  &$-0.8467(4)$  &$2\times10^6$  &$30$           &$>0.999$\\
3  &$-1.2530(2)$  &$2\times10^6$  &$10^2$         &$>0.999$\\
4  &$-1.3114(2)$  &$       10^6$  &$3\times10^2$  &$>0.999$\\
5  &$-1.3497(2)$  &$5\times10^5$  &$10^3$         &$>0.999$\\
6  &$-1.3661(2)$  &$2.5\times10^5$&$3\times10^3$  &$>0.999$\\
7  &$-1.3764(2)$  &$10^5$         &$7\times10^3$  &$>0.999$\\
8  &$-1.3820(3)$  &$5\times10^4$  &$1.6\times10^4$&$>0.999$\\
9  &$-1.3854(5)$  &$10^4$         &$10^4$         &$>0.999$\\
10 &$-1.3893(5)$  &$10^4$         &$10^4$         &$>0.999$\\
12 &$-1.3932(4)$  &$10^4$         &$10^4$         &$>0.999$\\
14 &$-1.3955(3)$  &$10^4$         &$10^4$         &$>0.999$\\
16 &$-1.3973(3)$  &$10^4$         &$2\times10^4$  &$0.996$\\
18 &$-1.3974(3)$  &$6\times10^3$  &$4.5\times10^4$&$0.996$\\
20 &$-1.3985(4)$  &$3\times10^3$  &$10^5$         &$0.996$\\
22 &$-1.3981(5)$  &$2\times10^3$  &$2.5\times10^5$&$0.996$\\
24 &$-1.3994(6)$  &$10^3$         &$5.5\times10^5$&$0.996$\\
26 &$-1.3992(12)$ &$2\times10^2$  &$1.2\times10^6$&$0.996$\\
28 &$-1.4001(15)$ &$10^2$         &$2.5\times10^6$&$0.996$\\
30 &$-1.3993(17)$ &$10^2$         &$5\times10^6$  &$0.996$\\
\hline
\end{tabular}
\end{table}
\begin{table}
\caption{\label{table3} Simulation parameters and GS energy per
spin for the 2D EAG model. }
\begin{tabular}{ccccc}
\hline
$L$&     $u_L$    &     $N_{\mathrm{s}}$     &      $t$      &$P_0$  \\
\hline
2  &$-1.0322(3)$  &$2\times10^6$  &$30$           & $>0.999$     \\
3  &$-1.2074(2)$  &$2\times10^6$  &$10^2$         & $>0.999$     \\
4  &$-1.2603(2)$  &$10^6$         &$3\times10^2$  & $>0.999$    \\
5  &$-1.2826(3)$  &$5\times10^5$  &$10^3$         & $>0.999$     \\
6  &$-1.2936(3)$  &$2.5\times10^5$&$3\times10^3$  & $>0.999$     \\
7  &$-1.3000(3)$  &$10^5$         &$7\times10^3$  & $>0.999$     \\
8  &$-1.3027(4)$  &$5\times10^4$  &$10^4$         & $>0.999$     \\
9  &$-1.3070(8)$  &$10^4$         &$10^4$         & $>0.999$     \\
10 &$-1.3072(7)$  &$10^4$         &$10^4$         &$0.998$     \\
12 &$-1.3103(6)$  &$10^4$         &$4\times10^4$  &$0.998$     \\
14 &$-1.3119(7)$  &$5\times10^3$  &$1.5\times10^5$&$0.998$     \\
16 &$-1.3119(8)$  &$3\times10^3$  &$5.3\times10^5$&$0.998$     \\
18 &$-1.3154(13)$ &$10^3$         &$1.7\times10^6$&$0.998$     \\
20 &$-1.3127(12)$ &$10^3$         &$2\times10^6$  &$0.99$     \\
22 &$-1.3144(22)$ &$2\times10^2$  &$2.7\times10^6$&$0.97$     \\
24 &$-1.3130(23)$ &$2\times10^2$  &$7.3\times10^6$&$0.97$     \\
26 &$-1.3146(26)$ &$10^2$         &$8.3\times10^6$&$0.9$     \\
\hline
\end{tabular}
\end{table}
\begin{table}
\caption{\label{table4} Simulation parameters and GS energy per
spin for the 3D EAB model. }
\begin{tabular}{ccccc}
\hline
$L$&     $u_L$    &     $N_{\mathrm{s}}$     &      $t$      &$P_0$  \\
\hline
2  &$-1.3473(4)$  &$10^6$         &$30$           &$>0.999$     \\
3  &$-1.6717(1)$  &$10^6$         &$10^2$         &$>0.999$     \\
4  &$-1.7375(1)$  &$5\times10^5$  &$5\times10^2$  &$>0.999$     \\
5  &$-1.7611(1)$  &$10^5$         &$2\times10^3$  &$>0.999$     \\
6  &$-1.7714(3)$  &$10^4$         &$2\times10^4$  &$>0.999$     \\
7  &$-1.7772(3)$  &$6\times10^3$  &$2\times10^5$  &$>0.999$     \\
8  &$-1.7800(3)$  &$4\times10^3$  &$7\times10^5$  &$0.997$\\
9  &$-1.7824(3)$  &$2\times10^3$  &$1.2\times10^6$&$0.99$ \\
10 &$-1.7830(3)$  &$2\times10^3$  &$2\times10^6$  &$0.97$ \\
12 &$-1.7849(8)$  &$10^2$         &$10^7$         &$0.93$ \\
14 &$-1.7858(7)$  &$10^2$         &$1.6\times10^7$&$0.73$ \\
\hline
\end{tabular}
\end{table}
\begin{table}
\caption{\label{table5} Simulation parameters and GS energy per
spin for the 3D EAG model. }
\begin{tabular}{ccccc}
\hline
$L$&     $u_L$    &$N_{\mathrm{s}}$&      $t$      &$P_0$  \\
\hline
2  &$-1.4360(3)$  &$10^6$         &$10^2$         & $>0.999$     \\
3  &$-1.6204(2)$  &$5\times10^5$  &$3\times10^2$  & $>0.999$     \\
4  &$-1.6660(2)$  &$2\times10^5$  &$1.5\times10^3$& $>0.999$     \\
5  &$-1.6824(3)$  &$6\times10^4$  &$10^4$         & $>0.999$     \\
6  &$-1.6891(4)$  &$2\times10^4$  &$5\times10^4$  & $>0.999$     \\
7  &$-1.6937(8)$  &$3\times10^3$  &$5\times10^5$  & $>0.999$     \\
8  &$-1.6955(6)$  &$3\times10^3$  &$10^6$         & 0.997 \\
9  &$-1.6966(7)$  &$2\times10^3$  &$1.3\times10^6$& 0.98  \\
10 &$-1.6981(7)$  &$1.3\times10^3$&$1.6\times10^6$& 0.90  \\
11 &$-1.6982(8)$  &$826$          &$4.1\times10^6$& 0.80  \\
\hline
\end{tabular}
\end{table}


\begin{thebibliography}{99}
\bibitem{Nota1} Nevertheless, for the B algorithm this property does not hold.
\end{thebibliography}
\end{document}